\documentclass{aa}  
\usepackage[colorlinks=true,linkcolor=blue,citecolor=blue,filecolor=blue,urlcolor=blue]{hyperref}
\usepackage{graphicx}
\usepackage{txfonts}
\usepackage{subcaption}         
\usepackage{lscape}             
\usepackage{placeins}           
                                
\AtBeginDocument{\let\linenumbers\nolinenumbers}

\begin{document}

   \title{Ultra-diffuse Galaxies in the Kilo-Degree Survey: Globular Cluster Systems in Cluster and Field Environments}
    \author{Hao Su
          \inst{1,2}
          \and
          Nicola R. Napolitano\inst{1,2,3}\fnmsep\thanks{Corresponding authors: Nicola R. Napolitano: nicolarosario.napolitano@unina.it; Rui Li: liruiww@gmail.com}
          \and
          Rui Li\inst{4}
          \and
          Crescenzo Tortora\inst{2}
          \and
          Fucheng Zhong\inst{5,2}
          \and
          Zhenping Yi\inst{6}
          \and
          Yang Liu\inst{1}
          }           
   \institute{Department of Physics “E. Pancini”, University of Naples Federico II, Via Cintia, 21, 80126 Naples, Italy
        \and
        INAF – Osservatorio Astronomico di Capodimonte, Salita Moiariello 16, I-80131 Napoli, Italy
        \and
        Scuola Superiore Meridionale, via Mezzocannone, 4, Napoli, Italy
        \and
        Institute for Astrophysics, School of Physics, Zhengzhou University, Zhengzhou, 450001, People’s Republic of China
        \and
        Purple Mountain Observatory, Chinese Academy of Sciences, Nanjing 210023, P. R. China
        \and
        School of Airspace Science and Engineering, Shandong University, 180 Wenhua Xilu, Weihai 264209, Shandong, China
    }
   \date{}
  \abstract  
   {\begin{abstract}
   
   Ultra-diffuse galaxies (UDGs) are an extreme population of low-surface-brightness systems whose formation and evolutionary pathways remain debated. Their globular cluster (GC) systems provide a powerful probe of early halo assembly and star-formation efficiency, yet systematic studies over wide survey areas are still limited. In this work, we investigate the GC systems of UDGs identified in the Kilo-Degree Survey (KiDS). Starting from a catalog of 545 high-quality UDG candidates previously identified using the deep-learning framework UDGnet, we developed a dedicated pipeline for the detection of GCs. 
   We identify 24 UDGs, including 14 systems associated with galaxy groups or clusters and 10 field systems, with $N_{\rm GC} \geq 3$, and use these to extrapolate the globular cluster luminosity function (GCLF) and estimate their distances and physical sizes. We find that group- and cluster-associated UDGs tend, on average, to exhibit somewhat higher GC numbers and specific frequencies than field UDGs at fixed luminosity and stellar mass, although the overlap between the two populations remains substantial. In particular, most systems with the highest specific frequencies ($S_N \gtrsim 40$) are associated with galaxy groups or clusters, while field UDGs generally populate the lower part of the overall UDG distribution without forming a clearly distinct sequence. At a given stellar mass, group- and cluster-associated UDGs more often occupy the upper envelope of the $N_{\rm GC}$--$M_\star$ relation. Using empirical GC--halo mass scaling relations, this corresponds to higher inferred ($M_h$) at fixed ($M_\star$), and therefore to lower stellar-to-halo mass ratios, with only a mild tendency for cluster UDGs to populate the higher-halo-mass end of the distribution. These results support a scenario in which UDGs formed their GC populations efficiently at early epochs, followed by suppressed or prematurely quenched field-star formation, and suggest that they may represent the high-GC-richness extension of the dwarf-galaxy population. The observed environmental trends may reflect differences in early assembly at a given stellar mass, together with potentially different quenching histories, preprocessing, infall times, and survival pathways, rather than a sharp present-day dichotomy between field and cluster systems.

\end{abstract}}


   \maketitle

\section{Introduction}
Since the identification of a large population of ultra-diffuse galaxies (UDGs) in the Coma cluster, these faint stellar systems have attracted considerable attention as probes of galaxy formation and evolution. UDGs are commonly defined as galaxies with extremely low central surface brightness ($\mu_{0,g} \geq 24$ mag arcsec$^{-2}$) and large effective radii ($R_e \geq 1.5$ kpc; \citealt{van2015forty}), often exhibiting stellar masses typical of dwarf galaxies but sizes comparable to those of much more massive systems. However, these empirical thresholds are somewhat arbitrary, as subsequent studies have shown that UDGs do not form a clearly distinct population from other low-surface-brightness (LSB) galaxies (\citealt{venhola2018fornax,venhola2022fornax,lim2020next,marleau2021ultra, van2022s}. Consequently, the operational definition of UDGs varies across studies, potentially affecting sample selection and the interpretation of their physical properties.

In the galaxy formation context, one of the central questions is whether UDGs represent ``failed'' galaxies residing in unusually massive dark-matter halos or extended systems formed within dwarf-mass halos.
In the former picture, UDGs reside in relatively massive dark-matter halos but have formed far fewer stars than expected for their halo mass.
In the latter picture, UDGs are intrinsically dwarf-mass galaxies, hosted by halos comparable to those of the LMC or by even lower-mass dwarf halos.
More broadly, UDGs' unusually extended and diffuse stellar distributions may arise through mechanisms such as high halo spin, internal feedback, tidal heating, environmental processing, and galaxy mergers, or a combination of these effects (e.g. \citealt{amorisco2016ultradiffuse, di2017nihao, rong2017universe, benavides2024large, forbes2025some}).
In particular, observations of morphologically disturbed UDGs suggest that tidal interactions and dwarf--dwarf mergers may contribute to their formation (\citealt{fielder2023disturbed, fielder2024all}). Recent hydrodynamical simulations further demonstrate that gas-rich dwarf mergers can produce UDG-like remnants hosting rich globular cluster systems (\citealt{tapia2026formation}).
As highlighted in the recent review by \citet{gannon2026dawes}, no single formation mechanism appears capable of explaining the full diversity of observed UDG properties.
This diversity of possible formation pathways motivates studies of UDG structural properties, stellar populations, and globular cluster systems across different environments.

In this context, globular cluster (GC) systems represent one of the most powerful diagnostics for probing the nature of UDGs. As they host some of the oldest stellar populations in galaxies, GCs are closely connected to the early assembly of their parent dark matter halos \citep{brodie2006extragalactic,harris2013catalog}. Their total numbers, spatial distributions, and scaling relations with host-galaxy properties therefore provide valuable constraints on halo mass, star-formation efficiency, and assembly history. In particular, the nearly linear relation between GC system mass, or equivalently GC number, and halo virial mass established for massive galaxies \citep{harris2017galactic,burkert2020high} has 
made the GCs an indirect tracers of dark matter halos in low-surface-brightness systems, where conventional dynamical measurements are 
hardly applicable
\citep[e.g.][]{van2016high,beasley2016overmassive,peng2016rich}. Hence, the total GC population, often normalized by host-galaxy luminosity through the specific frequency $S_N$, provides an effective means of constraining the halo masses of UDGs and discriminating among different formation scenarios.

Observational studies have applied GC-based diagnostics to UDGs across a wide range of environments, revealing substantial diversity in their globular cluster populations. Early studies of individual systems---such as VCC 1287 in Virgo and the Coma UDGs DF44 and DFX1 (\citealt[]{beasley2016overmassive, van2016high, van2017abundance})---suggested that some UDGs might host unusually rich GC systems, implying high halo-to-stellar mass ratios. Later work, however, revised the GC numbers for several of these extreme systems, including DF44 and DFX1, showing that their GC populations had likely been overestimated and that the case for massive dark matter halos in these galaxies is therefore less compelling (\citealt[]{saifollahi2022implications}). Statistical analyses of larger samples in Coma (\citealt{amorisco2018globular, lim2018globular, forbes2020globular}) further demonstrated that UDGs span a broad range of GC richness, with some systems lying above the typical $N_{\rm GC}-M_\star$ relation of classical dwarfs. Rather than supporting a picture in which UDGs are predominantly failed massive galaxies, these results point to a heterogeneous population in which some UDGs have relatively high halo-to-stellar mass ratios, while others are consistent with more ordinary dwarf-galaxy halos. Indeed, several studies have found UDGs with GC abundances similar to those of classical dwarfs (\citealt{amorisco2018globular, prole2019halo,muller2021dwarf,brook2021shallow}), supporting more modest halo masses and evolutionary pathways shaped by internal feedback, high halo spin, tidal heating, environmental processing, or a combination of these mechanisms. Quantitatively, some works report that UDGs host two to three times more GCs than dwarfs of the same stellar mass (\citealt{roman2017ultra, forbes2024ultra}), while others detect no systematic enhancement. Wide-field surveys such as MATLAS further reveal a broad continuum of GC richness, from GC-poor systems to galaxies with relatively rich GC populations (\citealt{lim2018globular,lim2020next}). Additional GC properties, including colors, luminosity functions, and radial distributions, appear broadly consistent with those of dwarf galaxies (\citealt{marleau2024dwarf}). In particular, UDGs typically exhibit GC half-number-radius ratios $R_{\mathrm{e,GC}}/R_{\mathrm{e,gal}} \simeq 0.7-1.7$ (\citealt{peng2016rich, montes2021disk, saifollahi2021number, saifollahi2022implications, gannon2022ultra}), comparable to those of dwarfs in the Local Group. Together, these studies highlight the power of GC systems to constrain UDG halo masses and formation histories, suggesting that UDGs form a heterogeneous population ranging from GC-poor dwarf-like systems to GC-enhanced galaxies with high halo-to-stellar mass ratios.

More recently, Euclid has opened a new avenue for studies of GC systems in nearby low-mass galaxies by combining high spatial resolution, wide-field coverage, and imaging from the broad visible $I_E$ band to the near-infrared $Y_EJ_EH_E$ bands. Using Euclid observations of the Fornax galaxy cluster, \citealt{saifollahi2025euclid} demonstrated the capability of Euclid to identify and characterize GC populations associated to dwarf galaxies as well as the intracluster medium. In the Perseus cluster, \citealt{saifollahi2025euclid1} also showed that low-surface-brightness dwarf galaxies, including UDGs, tend on average to host larger GC populations than more compact dwarfs at comparable stellar mass.

Despite this progress, several key questions remain open, particularly regarding how strongly GC richness depends on environment and how reliably GC populations can be used to infer dark-matter halo masses, especially in the dwarf-galaxy regime.
Although some of the most GC-rich UDGs have been identified in dense environments, the strength of this environmental dependence remains uncertain, partly because existing samples are heterogeneous and subject to selection effects.
For example, HST studies of gas-rich field UDGs have identified very few old GC candidates, broadly consistent with expectations for dwarfs of similar stellar mass and arguing against these systems being direct analogues of the GC-rich UDGs found in clusters (\citealt{jones2023gas}).
Similarly, studies of UDGs in group environments indicate that most systems have GC abundances comparable to those of normal dwarfs, although a few exhibit enhanced GC populations (\citealt{somalwar2020hyper}).
Whether UDGs systematically differ from normal dwarf galaxies in their GC populations at fixed stellar mass, and how such differences depend on environment, therefore remains an open question.

An additional complication is that the present-day environment does not necessarily reflect the full evolutionary history of a UDG.
Stellar-population studies suggest that some UDGs in the Coma cluster may have been accreted relatively recently, implying that they have experienced less prolonged cluster processing than long-term cluster members \citep{alabi2018origins}.
By contrast, phase-space studies of UDGs in the Hydra I cluster suggest earlier infall, with no clear evidence for a universal connection between GC richness and accretion time \citep{forbes2023ultra}.
These findings indicate that observed environmental trends may reflect a combination of preprocessing, infall time, orbital history, and subsequent interactions within the host system.
Some cluster UDGs exhibit high GC specific frequencies ($S_N$), although their GC populations are diverse \citep[e.g.][]{prole2019observational, kado2022ultra}.
Field UDGs, by contrast, span a broad range of GC richness and inferred halo masses and may follow alternative evolutionary pathways, including extended star formation or secular expansion in high-spin dwarf haloes \citep{brook2021shallow, chan2018origin, benavides2023origin}.
These findings suggest that GC richness may be influenced by both formation history and environmental processes, highlighting the importance of investigating GC populations across different environments.

A key observational challenge in investigating these environmental trends is obtaining reliable and homogeneous GC measurements across large UDG samples.
The selection of high-purity GC samples for accurate number counts critically depends on controlling contamination from foreground stars and unresolved background galaxies.
Combining multi-band optical imaging with near-infrared photometry can provide additional colour information to improve the discrimination between genuine GCs and contaminants (e.g. \citealt{munoz2014next}).
This complementarity is particularly relevant for space-based observations. For example, \citet{saifollahi2025euclid} demonstrated the capabilities of Euclid for GC detection and highlighted the value of additional ground-based optical imaging for refining photometrically selected GC samples.
Wide-field optical surveys, such as the Hyper Suprime-Cam Subaru Strategic Program (HSC-SSP; \citealt{aihara2018hyper}) and the Kilo-Degree Survey (KiDS; \citealt{wright2024fifth}), provide high-quality multi-band imaging suitable for systematic GC searches and complementary analyses with space-based observations from Euclid and Roman (\citealt{schlieder2024survey}).
Such datasets offer an opportunity to build more homogeneous GC catalogues and investigate the connection between GC populations and the environments of UDGs.

In this work, we make use of a new catalog of UDG candidates identified in the KiDS DR5 survey \citep{su2026ultra}, to investigate the globular cluster systems over a wide range of environments in a homogeneous and statistical way. Using a dedicated pipeline, we detect and characterize GC candidates associated with each UDG, allowing us to quantify both the richness and spatial properties of their GC systems across a wide range of stellar masses and environments. Our analysis focuses in particular on GC-rich UDGs and explores how their properties vary as a function of environment, with special emphasis on the comparison between field and cluster populations. By comparing our results with previous studies in both cluster and group environments, we aim to place UDGs within a coherent framework that links their GC systems to their dark-matter halos and evolutionary histories, and to constrain their halo masses and formation pathways.

This paper is organized as follows. In Sect.~\ref{data} we describe the KiDS--DR5 data and the parent high-quality UDG catalog. In Sect.~\ref{method} we present our GC-detection pipeline, including validation tests, and outline the construction of the final GC-rich UDG sample. In Sect.~\ref{analysis} we present our main results on the spatial distribution, environmental association, and GC scaling relations. Finally, in Sect.~\ref{conclusion} we summarize our results and conclusions.

\section{Data}
\label{data}
The Kilo-Degree Survey (KiDS) is a deep optical imaging programme carried out with the 2.6-m VLT Survey Telescope at ESO’s Cerro Paranal Observatory, using the wide-field OmegaCAM instrument to map the sky in the \textit{ugri} bands over a footprint of approximately $1350~\mathrm{deg}^2$. As part of its design for precision weak-lensing measurements, KiDS employs a dedicated observing strategy, including strict seeing constraints, rotational dithering, and nightly photometric calibration, to ensure uniform image quality across all survey tiles \citep{de2015first}. The resulting data achieve a median $r$-band point-spread function (PSF) full width at half maximum (FWHM) of about $0.7'' $, and $5\sigma$ point-source depths of $u \simeq 24.3$, $g \simeq 25.2$, $r \simeq 24.8$, and $i \simeq 23.7$ mag \citep{kuijken2019fourth}. The fifth and last data release, KiDS-DR5 \citep{wright2024fifth}, presents the complete survey area with improved photometric homogenisation and updated astrometric solutions, yet incorporating matched near-infrared imaging from the VIKING survey \citep{edge2013vista,venemans2015first}. This finally makes the KiDS-DR5 the widest and coherent nine-band optical--near-infrared dataset yet available to the community. Processing of DR5 images is performed within the \textsc{ARTEMIS} pipeline, which provides refined background modeling and enhanced control of large-scale systematics \citep{wright2024fifth}. These characteristics, together with the dark-sky conditions at Paranal, allow KiDS-DR5 to reach a $3\sigma$ limiting surface brightness of $\mu_{r,\mathrm{lim}} \simeq 26.4~\mathrm{mag~arcsec^{-2}}$, estimated over $10'' \times 10''$ regions. This surface-brightness depth enables the detection and characterization of LSBGs, including UDGs \citep{roy2018evolution, kelvin2018galaxy}.

In our previous work, we identified a catalog of UDG candidates using the deep-learning-based object detection framework UDGnet \citep{su2026ultra}. This catalog was constructed through a large-scale iterative detection method followed by multiple rounds of visual inspection, resulting in a sample of 545 High-Quality (HQ) UDG candidates spanning the full KiDS footprint. In the present analysis, this HQ UDG catalog serves as the parent sample for the identification of associated globular cluster (GC) candidates. A dedicated GC-detection pipeline is applied uniformly to all HQ UDGs. 

\begin{figure*}[ht!]
    \centering
    \includegraphics[width=0.9\textwidth]{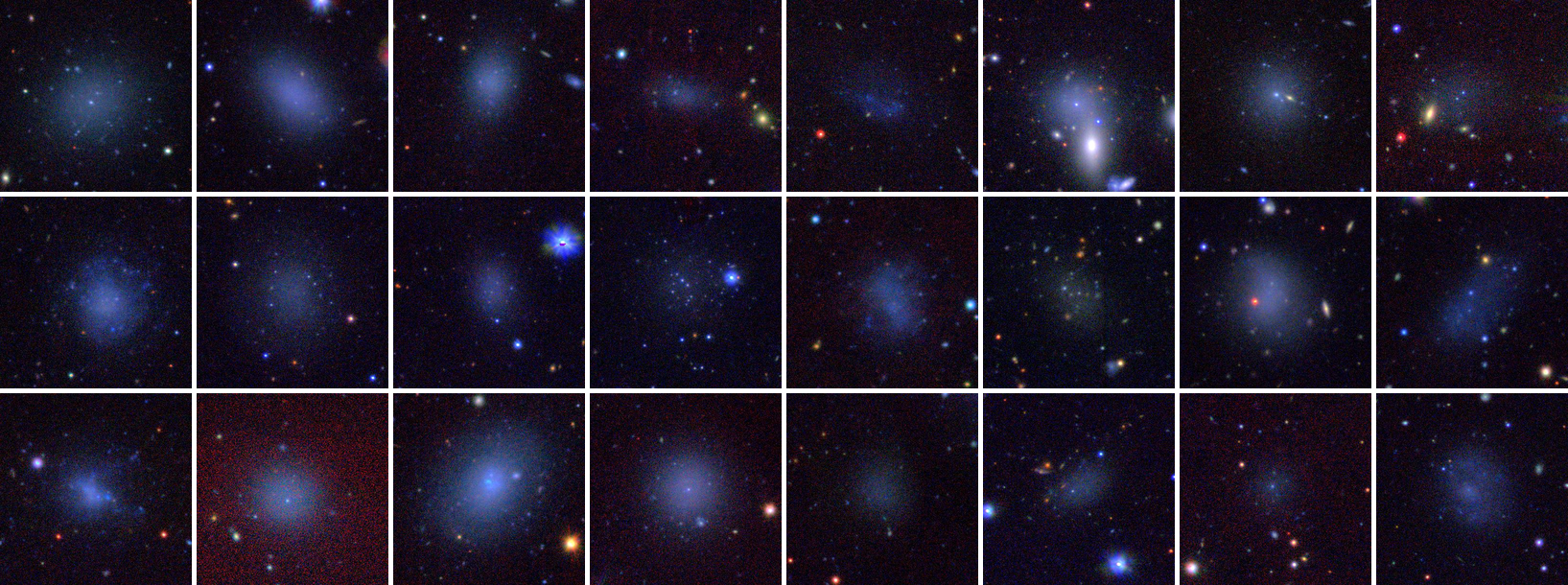}
    \caption{Images of the 24 GC-rich UDG candidates analysed in this work, selected with  $\langle \mu_{e}\rangle_r>24$ mag/arcsec$^2$ and $ R_e\geq 1.3$ kpc. The first 14 panels show UDGs associated with galaxy groups or clusters, while the remaining 10 panels correspond to field UDGs. The size of each cutout is $80'' \times 80''$.}
    \label{fig:UDG_images}
\end{figure*}

Following previous studies, we select a subsample of UDGs hosting at least three GC candidates within $2R_e$, which we refer to as \emph{GC-rich} systems \citep{saifollahi2022implications, jones2023gas, marleau2024dwarf, saifollahi2025euclid1,marleau2025euclid}.The choice of this threshold and the resulting sample selection are discussed in detail in Section~\ref{sec:Application to the HQ KiDS UDG catalog}. For each GC-rich UDG, we construct the globular cluster luminosity function (GCLF) and, for systems with sufficient GC statistics, derive distance estimates and corresponding physical effective radii. This procedure allows us to identify UDGs reliably located within the distance sensitivity range of KiDS and to assemble a refined sample suitable for subsequent structural and environmental analyses. The final sample consists of 24 UDGs, including 10 field systems and 14 UDGs associated with galaxy groups or clusters. The images of these 24 UDGs are shown in Fig.\ref{fig:UDG_images}.

\section{Method}
\label{method}
\subsection{Detection of GCs in KiDS UDGs}

Given that the $r$ band provides the best image quality in KiDS--DR5, with a median seeing of approximately $0.7''$, we perform the primary detection of GC candidates on the $r$-band images. The nominal $5\sigma$ point-source limiting magnitude in this band, $r \simeq 24.8$ mag, would in principle allow the GCLF turnover to be reached out to distances of order $\sim 28.8$ Mpc. However, this value corresponds to the faint-end detection limit, where completeness is already significantly reduced. Moreover, the final GC-candidate sample is selected using multi-band photometry and colour cuts, so its effective completeness is not set by the $r$ band alone, but also by the shallower and lower-quality bands, in particular the $g$ band.

We also tested an independent extraction of compact-source candidates from the $g$-band images. Owing to the poorer seeing and lower effective depth of the $g$ band, this procedure tends to miss part of the compact-source population detected in the $r$ band and, for most UDG fields, did not reveal additional reliable GC candidates. We therefore adopt the $r$ band for source detection and use the other bands for photometric color selection. Adopting a more conservative effective completeness magnitude of $r \sim 24.7$ mag, and considering the additional limitation imposed by the color selection, we estimate that the GCLF turnover can be sampled with reasonable completeness only for the nearest systems, approximately within $\sim 24$ Mpc. At larger distances, the observed GC population increasingly samples only the bright side of the luminosity function; therefore, the derived GC numbers should be interpreted as completeness-corrected estimates based on a partially sampled GCLF, rather than as direct complete counts.

For each UDG, we start from the KiDS--DR5 $r$-band image and the corresponding bounding box provided by UDGnet, which provides the UDG center and extent in normalized image coordinates. These coordinates are converted into pixel units, and we extract a fixed-size cutout of $400 \times 400$ pixels ($80'' \times 80''$), centred on the UDG. Within this cutout, we adopt a circular GC search region with a projected galactocentric radius of $2R_{\rm e}$ \citep{saifollahi2022implications, marleau2024dwarf, saifollahi2025euclid1}. This allows us to search for GC candidates within twice the effective radius of the UDG while retaining the surrounding area for background estimation. Such a cropping strategy allows us to focus on the region most relevant for GC detection while avoiding the unnecessary inclusion of distant background structures. Obvious artifacts or saturated features within the cutout are mitigated by replacing pixels exceeding 100 times the mean cutout value with the local median, thereby preventing them from biasing the subsequent background modeling.

We then model and subtract the large-scale background using the \texttt{Background2D} algorithm with a median estimator. The background is computed on a grid whose box size is chosen adaptively as

\begin{equation}
N_{\mathrm{box}}=\operatorname{clip}\left(\frac{\min(N_x,N_y)}{6},\,50,\,128\right)
\end{equation}
where $N_{\mathrm{box}}$ is the background-mesh size in pixels, and $N_x$ and $N_y$ are the dimensions of the cutout in pixels. This choice avoids both oversmoothing caused by excessively large boxes and undersampling resulting from boxes that are too small. The resulting two-dimensional background model is then subtracted from the cutout to produce a residual image optimized for compact-source detection.

Compact sources are detected in the residual image using \texttt{DAOStarFinder} \citep{robitaille2013astropy}\footnote{\url{https://github.com/astropy/photutils/tree/main/photutils/detection}}. We first calculate the sigma-clipped background statistics (mean, median, and standard deviation) using a $3\sigma$ clipping threshold and adopt a detection threshold of $5\sigma$ above the local background noise. The PSF scale is determined from the median KiDS $r$-band seeing of $0.7''$. Given the KiDS pixel scale of $0.2\,\mathrm{arcsec\,pixel^{-1}}$, this seeing corresponds to a PSF FWHM of $3.5$ pixels. Assuming a Gaussian PSF, the corresponding standard deviation is $\sigma_{\rm PSF} \simeq 1.48$ pixels, yielding a $3\sigma_{\rm PSF}$ scale of approximately $4.44$ pixels.

Thus, a $3\sigma$ aperture corresponds to approximately 4.5 pixels, capturing $\sim 98.9\%$ of the flux of a Gaussian-like PSF. However, since the $0.7''$ KiDS $r$-band seeing represents an average value and varies across survey tiles, we adopt a more conservative choice and round the $3\sigma$ aperture radius up to 5 pixels \citep[e.g.][]{kuijken2019fourth, wright2024fifth}. This radius is adopted for the subsequent aperture photometry, enclosing most of the point-source flux while limiting background contamination. Based on this PSF model, we retain only sources with measured $\mathrm{FWHM} < 2 \times \mathrm{FWHM}_{\rm seeing}$, reducing contamination from extended background galaxies and diffuse structures.

For all remaining compact sources, we perform aperture photometry on the background-subtracted image. We adopt a circular aperture with a radius of 5 pixels ($1.0''$) to measure the source flux, and estimate the local background using a circular annulus with inner and outer radii of 8 and 12 pixels ($1.6''$--$2.4''$), respectively. For each source, the median pixel value within the annulus is taken as the local background level, multiplied by the aperture area, and subtracted from the raw aperture sum to obtain the net source flux. Instrumental magnitudes are then computed as
\begin{equation}
\label{eq1}
m_{\rm inst} = -2.5 \log_{10}(F_{\rm net}),
\end{equation}
where $F_{\rm net}$ is the background-subtracted flux measured within the aperture.

To correct for flux losses due to the finite aperture size, we derive an aperture correction using bright, isolated stars within the same cutout. For each reference star, we measure background-subtracted fluxes within a small aperture of radius 5 pixels and a larger aperture of radius 8 pixels \citep{prole2019halo, cantiello2020fornax, marleau2024dwarf}. The corresponding magnitude difference is computed as
\begin{equation}
\Delta m = -2.5 \log_{10}
\left(\frac{F_{\rm 5\,px}}{F_{\rm 8\,px}}\right).
\end{equation}
The median value over all suitable stars is adopted as the aperture correction for that cutout.
For an ideal Gaussian PSF with a FWHM of $0.7''$, the correction between the 5- and 8-pixel apertures is approximately 0.004 mag. However, the empirical correction accounts for deviations from the Gaussian profile and PSF variations across the survey tiles. This correction is then subtracted from the instrumental magnitudes of all GC candidates ($m_{\rm corr}=m_{\rm inst}-\Delta m$).

The positions of the detected candidates are subsequently transformed back to the full-image coordinate system, allowing us to compute the projected galactocentric distance of each source relative to the UDG center. Distances are expressed both in pixels and in arcseconds. To estimate the level of foreground and background contamination, we define up to four control regions, each having the same size as the corresponding UDG cutout. These regions are placed along the four cardinal directions, at offsets of 1.5 times the cutout dimensions, provided that they remain fully within the image boundaries. In each control region, we repeat the same background-subtraction and source-detection procedure and count compact sources selected using the same morphological criteria.

The mean contamination estimate, $\langle N_{\rm ctrl}\rangle$, is calculated by averaging the detected source counts across the available control regions. For the final sample of 24 UDG candidates, the rounded values of $\langle N_{\rm ctrl}\rangle$ are 0 for 15 systems, 1 for 8 systems, and 2 for only 1 system. The mean contamination can be exactly zero when no sources satisfying our selection criteria are detected in any of the control regions. These results indicate that the background contamination is generally low and has a limited impact on the corrected GC counts. We emphasize that the unrounded values of $\langle N_{\rm ctrl}\rangle$ are used for background subtraction, while the resulting GC counts are rounded to the nearest integer only for presentation.

The standard deviation of the compact-source counts among the control regions, $s_{\rm ctrl}$, is used to characterize the local spatial variation in the foreground and background contamination. The uncertainty in the contamination-corrected GC richness is then estimated by combining the uncertainty in the mean contamination estimate in quadrature with the Poisson uncertainty of the raw candidate count:
\begin{equation}
\sigma_{N_{\rm GC}}=
\sqrt{N_{\rm raw}+\frac{s_{\rm ctrl}^{2}}{n_{\rm ctrl}}}.
\end{equation}
where $n_{\rm ctrl}$ is the number of valid control regions, which can vary depending on the image boundaries, with a maximum of four.

After completing the GC detection in the (r) band, we use the (r)-band positions of all compact candidates to perform aperture photometry in the (g) and (i) bands. This ensures that colors are measured consistently at identical positions, avoiding mismatches that could arise from independent detections in multiple bands under varying PSF conditions. We then apply a color cut to isolate likely GC candidates. Previous studies of GC populations in UDGs (e.g. \citealt{van2016high, amorisco2018globular, prole2019halo}) typically adopt a color range of $0.7 < (g-i) < 1.3$. To avoid excluding genuine GCs because of the photometric uncertainties in the KiDS images, we adopt a slightly expanded color interval of $0.6 < (g-i) < 1.4$. Although near-infrared photometry can provide additional leverage for separating GCs from contaminating sources, we do not include the overlapping VIKING (ZYJH$K_{s}$) imaging in the present analysis. Our detection and photometric procedure is optimized for the comparatively high image quality of the KiDS optical data, which is particularly important for sources that are unresolved or only marginally resolved at the distance of our UDG candidates. The generally broader PSF of the VIKING imaging would require a dedicated treatment of PSF matching, aperture corrections, and completeness before optical--NIR colors could be incorporated consistently into the present pipeline. We therefore restrict the current analysis to the homogeneous KiDS (gri) imaging. Nevertheless, the addition of VIKING near-infrared photometry could further improve the discrimination between genuine GCs and contaminants, and we plan to investigate its effectiveness in future work. 

In addition, the adopted region of color space is not unique to GCs and may contain foreground Milky Way stars, particularly cool dwarf stars, as well as unresolved or marginally resolved background galaxies whose colors overlap those of old GC populations. Photometric scatter can introduce additional contaminants near the boundaries of the adopted interval. We reduce stellar and galaxy contamination through our compact-source and FWHM selection criteria, while the remaining contamination is estimated statistically using the corresponding control regions. Sources falling outside the adopted color range are excluded from the final GC sample.

Finally, for each UDG we construct a radial surface-density profile of compact sources by binning GC candidates into concentric annuli centered on the UDG and normalizing by the annular area \citep{marleau2024dwarf}. This profile is used to verify the presence of a centrally concentrated GC system and to further assess the contrast between the UDG-associated GC population and the background level inferred from the control regions.

\subsection{Validation of the GC-detection pipeline}

\begin{figure}[htbp]
  \centering
\hspace{-0.5cm}
    \includegraphics[width=0.48\textwidth]{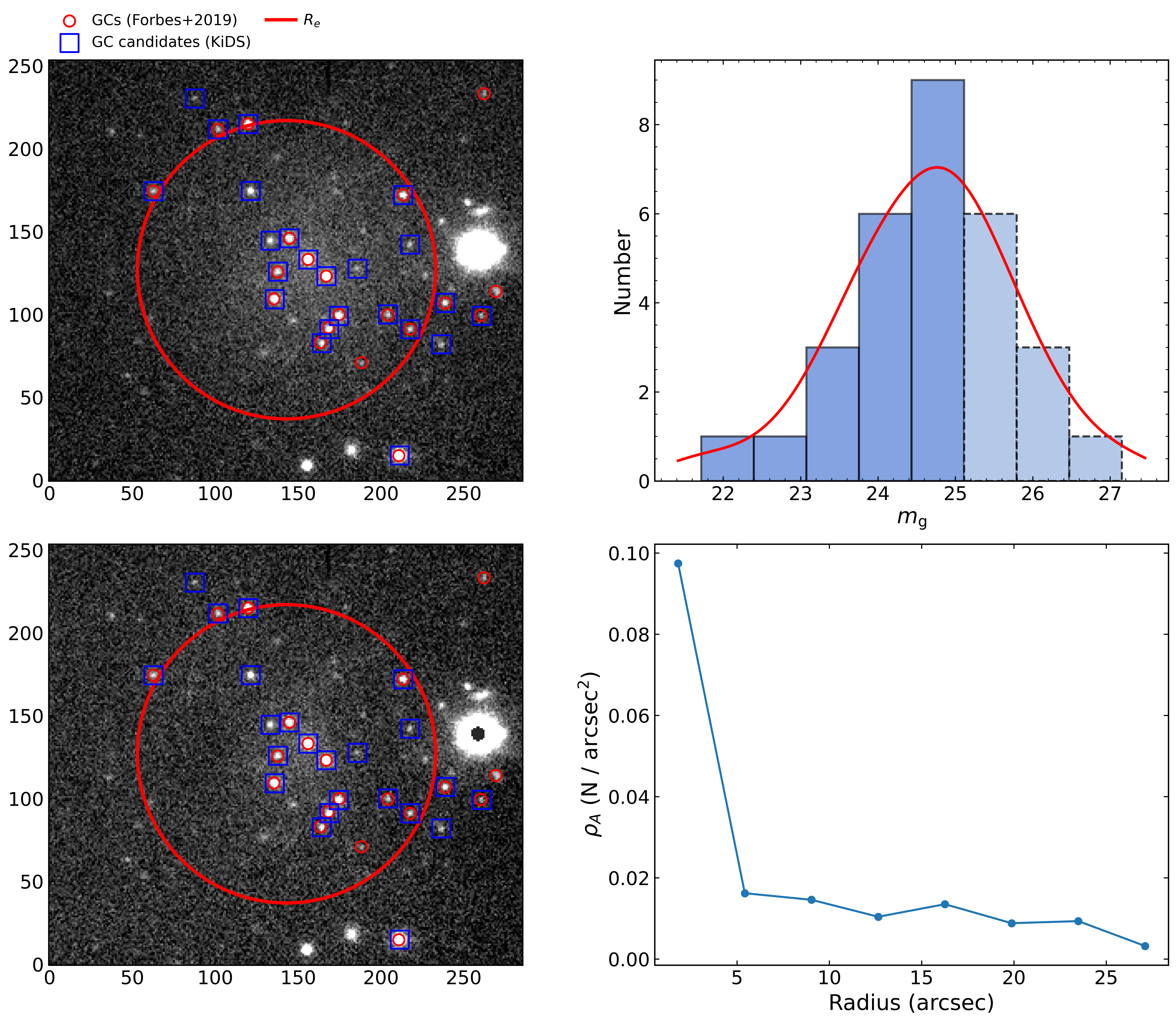}
    \caption{Validation of the GC-detection pipeline using a UDG in the NGC~5846.
    \emph{Upper left:} spatial distribution of GC candidates overlaid on the original KiDS-DR5 \emph{r}-band image.
    \emph{Upper right:} GCLF of the detected candidates, where the dashed component represents the estimated contribution from undetected faint GCs inferred from the assumed Gaussian shape of the GCLF.
    \emph{Lower left:} spatial distribution of GC candidates identified after processing the image with our GC-detection pipeline.
    \emph{Lower right:} radial surface-density profile of the GC system derived from the pipeline-selected candidates.
    }
    \label{fig:gc_validation_ngc5846}
\end{figure}

To validate the performance and reliability of our GC-detection pipeline, we first apply it to NGC~5846\_UDG1, a previously reported UDG in the NGC~5846 group for which a globular cluster census is available in the literature. We use the KiDS-DR5 \emph{r}-band image of this UDG as input and process it through our pipeline, identifying a total of 23 GC candidates associated with this galaxy. Among these, 17 sources are found to be in common with the compact-source catalog reported by \citet{forbes2019ultra}. This is consistent with their detection of $\sim 20$ compact sources, of which $\sim 3$ are expected to be background contaminants, corresponding to approximately 17 likely GC candidates. This agreement supports the ability of our pipeline to recover the bright GC-candidate population, although the 23 detected candidates should not be directly interpreted as the completeness-corrected total GC population. The spatial distribution of the detected GCs, together with the corresponding radial surface-density profile, is shown in Fig.~\ref{fig:gc_validation_ngc5846}.

We note that a range of total GC population estimates has been reported in the literature for this UDG. Based on single-orbit \emph{HST}/ACS imaging, \citet{muller2021dwarf} estimated $N_{\rm GC}=26\pm6$. From two-orbit \emph{HST}/WFC3 imaging, \citet{danieli2022ngc} derived $N_{\rm GC}=54\pm9$, assuming a distance of 26.5~Mpc. More recently, \citet{guerra2026mysterious} re-analyzed the GC system using existing \emph{HST} data together with deep $u$-band imaging and reported $N_{\rm GC}=33\pm3$. However, \citet{forbes2026revisiting} revisited these studies and obtained consistent total GC populations of $48^{+12}_{-8}$ and $50^{+4}_{-8}$, respectively, by counting GCs brighter than the GCLF turnover magnitude and assuming a symmetric GCLF. Their analysis supports a total population of approximately 50 GCs.

It is important to distinguish these literature estimates, which refer to the total GC population, from the 23 candidates directly detected by our KiDS-based pipeline. The estimated $r$-band GCLF turnover magnitude is approximately $m_{r}\simeq24.75$ mag, close to the median KiDS-DR5 $r$-band $5\sigma$ limiting magnitude of $\sim24.8$ mag. This suggests that our observations primarily probe the brighter portion of the GCLF. Assuming that our detected candidates approximately represent the brighter half of the GCLF, a simple factor-of-two scaling would imply a total population of $\sim46$ GCs, broadly consistent with the latest estimate by \citet{forbes2026revisiting}. However, this scaling is only illustrative, as the inferred total population may also be affected by photometric completeness, contamination, and the adopted GC search radius.

Furthermore, the lower spatial resolution and single-band selection of the KiDS data limit our ability to identify faint GCs and distinguish them from contaminants compared with the deeper, multi-band \emph{HST} observations. Nevertheless, the agreement in the detected bright GC population, together with the approximate consistency with recent total population estimates, provides a useful validation of our pipeline within the limitations of the KiDS data.

\begin{figure*}[ht!]
    \centering
    \includegraphics[width=0.98\textwidth]{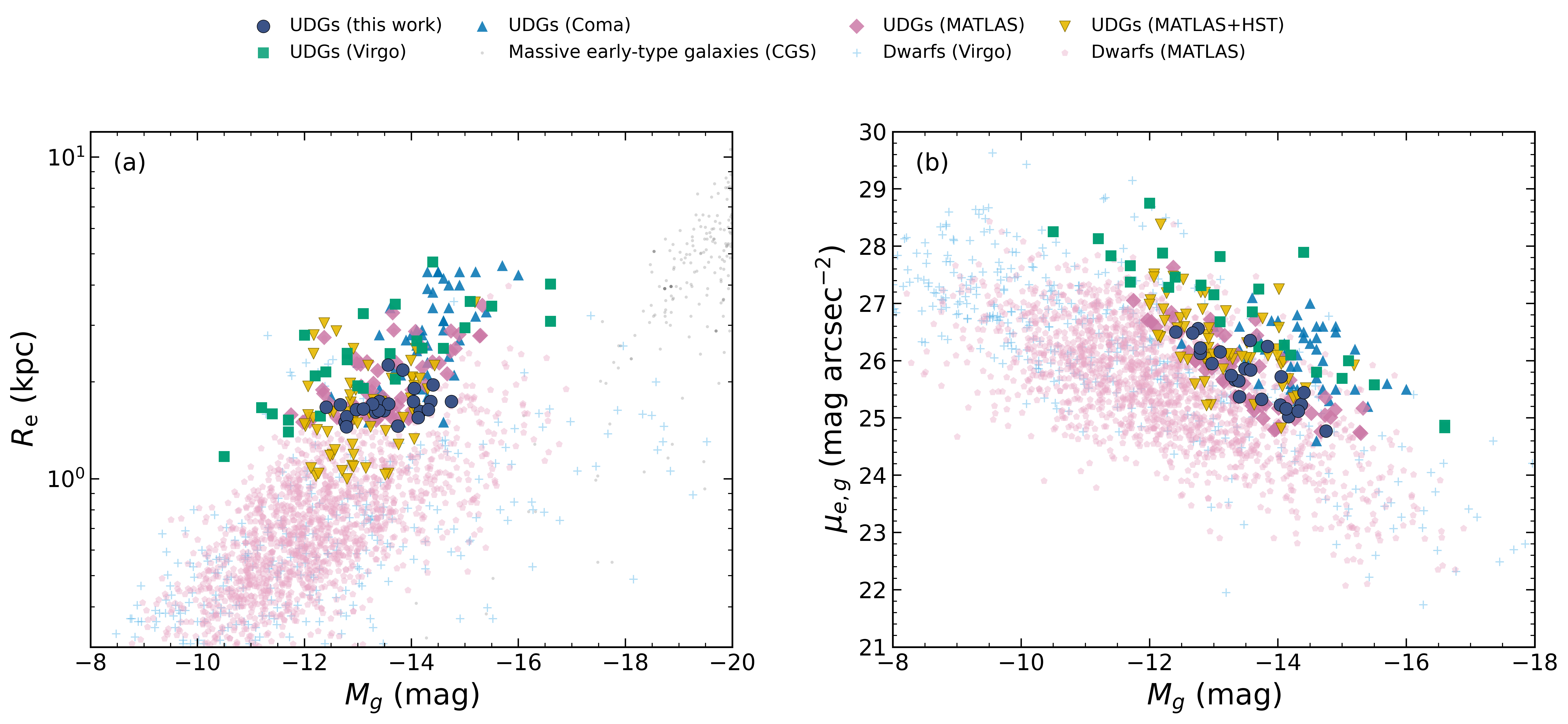}
    \caption{Distributions of key structural and photometric parameters for the final sample of GC-rich UDG candidates in KiDS, selected to have mean effective surface brightnesses fainter than $\langle\mu_e\rangle_r>24$ mag~arcsec$^{-2}$ and physical effective radii $R_{\rm e}>1.3$~kpc, compared with literature samples of UDGs, dwarf galaxies, and massive early-type galaxies. For the KiDS sample, the quoted physical effective radii have been corrected by a factor of 1.038. The left and right panels show the relations between absolute $g$-band magnitude and effective radius ($M_g$--$R_{\rm e}$) and between absolute magnitude and effective surface brightness ($M_g$--$\mu_{e,g}$), respectively. The effective-radius axis is logarithmic. Comparison objects outside the displayed ranges are omitted for clarity. Dark-blue circles denote KiDS UDG candidates from this work. Blue upward-pointing triangles represent Coma UDGs from \citet{van2015forty}. Pink diamonds correspond to MATLAS UDGs from \citet{marleau2021ultra}, while yellow downward-pointing triangles show MATLAS+HST UDGs from \citet{marleau2024dwarf}. Green squares indicate Virgo UDGs from \citet{lim2020next}. Grey points show massive early-type galaxies from the Carnegie--Irvine Galaxy Survey \citep{ho2011carnegie}, included only in the left panel. Light-blue plus symbols denote Virgo dwarf galaxies from \citet{ferrarese2020next}, and light-pink pentagons represent MATLAS dwarf galaxies from \citet{habas2020newly}.
    }
    \label{fig:parameter_distribution}
\end{figure*}
We further compare the photometry of the 17 GCs in common between our GC sample and that presented by \citet{forbes2019ultra}. The difference in the $g$-band magnitudes is found to be
\[
\Delta g = g_{\rm this\ work} - g_{\rm Forbes} = -0.09 \pm 0.22~{\rm mag},
\]
with a distribution consistent with random scatter and no evidence for systematic offsets. We attribute the residual dispersion primarily to small positional differences between the two catalogs, which can lead to minor variations in aperture centering and local background estimation, particularly for faint sources near the detection limit.

As an additional consistency check, we construct the GCLF of NGC~5846\_UDG1 using the $g$-band photometry; the result is shown in Fig. \ref{fig:gc_validation_ngc5846}. Adopting an absolute GCLF turnover magnitude of $M_g = -7.2$, the distance is estimated using

\begin{equation}
D({\rm Mpc}) = 10^{(m - M_{g} - 25)/5}.
\label{distance}
\end{equation}

Using this approach, we obtain a distance of $D \simeq 24.55$~Mpc for the UDG, in excellent agreement with the literature value of $D = 24.89$~Mpc reported by \citet{forbes2019ultra}. This close consistency provides an independent validation of both our GC photometry and our overall detection methodology.

In addition to photometric consistency checks, we also examined the literature for independent spectroscopic and kinematic confirmations of globular clusters associated with this system. In particular, the kinematic study of the GC system in NGC~5846~UDG1 by \citet{haacke2025investigating} provides radial-velocity measurements for a sample of 20 spectroscopically confirmed GCs associated with this UDG. 

We cross-matched our GC candidate catalog with their kinematically confirmed sample using positional matching and found that 80\% (16) of our detected GCs are independently confirmed by their kinematic analysis. This high level of overlap provides an additional, independent validation of the reliability of our GC-detection pipeline in identifying bona fide globular clusters associated with UDGs, despite the more limited spatial resolution and depth of the ground-based KiDS imaging compared to spectroscopic and \emph{HST}-based studies.

The remaining four kinematically confirmed GCs that are not recovered in our catalog can be naturally explained by observational limitations. Three of these objects have magnitudes close to the KiDS detection limit, where completeness rapidly declines. The last one exhibits marginally resolved or diffuse morphologies in the KiDS images, making them difficult to robustly classify as point-like sources. These cases are most likely missed due to the limited spatial resolution of ground-based imaging, rather than deficiencies in the detection strategy itself.

We further validate our GC-detection pipeline by applying it to UDGs in the Fornax cluster. Within the high-quality (HQ) UDG catalog presented in this work, two sources have previously been reported by \citet{venhola2017fornax} and subsequently analyzed by \citet{prole2019halo}, who investigated the globular cluster systems of Fornax low-surface-brightness galaxies using deep VST imaging.

In their study, GC candidates were identified using a catalog-based statistical approach based on morphological, color, and magnitude selections. Total GC populations were estimated after statistical background subtraction, but no sky positions or radial surface-density profiles of individual GCs were provided. As a result, a direct source-by-source comparison is not possible, and the validation can only be carried out at the level of total GC numbers.

\begin{table*}[t]
\centering
\caption{Main structural and photometric parameters of the 24 high-quality UDG candidates in KiDS.}
\label{table1}
\setlength{\tabcolsep}{2pt}
\renewcommand{\arraystretch}{1.09}
\scriptsize
\resizebox{\textwidth}{!}{%
\begin{tabular}{lcccclcccccc}
\hline\hline
Name & R.A. & Dec. & $N_{\rm GC}$ & $D$ & Environment & $R_{\rm e}$ & $\langle\mu_{\rm e}\rangle_r$ & $S_N$ & $\log_{10}(M_\star/M_\odot)$ & $\log_{10}(M_h/M_\odot)$ & $M_{\rm GCS}/M_\star$ \\
 & (deg) & (deg) & & (Mpc) & & (kpc) & (mag\,arcsec$^{-2}$) & & & & (\%) \\
\hline
    KiDS\_7.98\_-33.27 & 7.9805 & -33.2661 & $18\pm4$ & 25.10 & Field & 1.86 & 24.53 & 19.87 & 8.50 & 10.79 & 0.57 \\
    KiDS\_8.78\_-28.05 & 8.7793 & -28.0503 & $14\pm4$ & 26.40 & Field & 2.08 & 25.66 & 39.88 & 7.78 & 10.68 & 2.32 \\
    KiDS\_10.55\_-31.65 & 10.5496 & -31.6502 & $5\pm3$ & 21.40 & Field & 1.36 & 24.60 & 11.21 & 7.95 & 10.24 & 0.56 \\
    KiDS\_23.44\_-29.69 & 23.4424 & -29.6900 & $8\pm3$ & 22.40 & Field & 1.45 & 24.28 & 14.56 & 8.33 & 10.44 & 0.37 \\
    KiDS\_42.95\_-30.11 & 42.95023 & -30.1091 & $12\pm4$ & 20.40 & Field & 1.64 & 24.12 & 10.14 & 8.47 & 10.62 & 0.41 \\
    KiDS\_50.81\_-29.05 & 50.8052 & -29.0498 & $20\pm5$ & 24.20 & Field & 1.64 & 24.43 & 30.45 & 8.32 & 10.84 & 0.96 \\
    KiDS\_51.72\_-35.52 & 51.7249 & -35.5178 & $26\pm5$ & 21.90 & Fornax & 1.81 & 24.79 & 39.47 & 8.48 & 10.95 & 0.86 \\
    KiDS\_52.12\_-33.73 & 52.1181 & -33.7348 & $11\pm3$ & 22.80 & Fornax & 1.54 & 24.35 & 14.27 & 8.34 & 10.58 & 0.50 \\
    KiDS\_53.93\_-35.34$^{***}$ & 53.9281 & -35.5142 & $8\pm3$ & 19.05 & Fornax & 1.46 & 25.35 & 38.29 & 7.98 & 10.44 & 0.84 \\
    KiDS\_53.98\_-34.83$^{***}$ & 53.98202 & -34.8283 & $7\pm3$ & 19.05 & Fornax & 1.61 & 25.21 & 61.79 & 7.40 & 10.38 & 2.79 \\
    KiDS\_170.58\_3.90 & 170.5836 & 3.9008 & $5\pm2$ & 25.70 & NGC~3640 & 1.55 & 25.73 & 21.28 & 7.46 & 10.24 & 1.73 \\
    KiDS\_183.85\_3.31 & 183.8481 & 3.3083 & $6\pm2$ & 25.10 & Field & 1.40 & 25.49 & 27.82 & 8.07 & 10.32 & 0.51 \\
    KiDS\_189.95\_0.04 & 189.9486 & 0.0408 & $11\pm3$ & 24.20 & Field & 1.54 & 25.25 & 46.71 & 7.83 & 10.58 & 1.63 \\
    KiDS\_190.47\_2.74 & 190.4660 & 2.7446 & $13\pm4$ & 19.00 & NGC 4636 & 1.52 & 24.06 & 17.74 & 8.57 & 10.65 & 0.35 \\
    KiDS\_191.23\_2.30 & 191.2318 & 2.2959 & $17\pm4$ & 19.00 & NGC 4636 & 1.51 & 24.67 & 48.39 & 8.29 & 10.77 & 0.87 \\
    KiDS\_192.51\_3.22 & 192.5062 & 3.2154 & $10\pm3$ & 19.00 & NGC 4636 & 1.60 & 25.40 & 49.89 & 8.15 & 10.56 & 0.71 \\
    KiDS\_213.42\_-2.19 & 213.4208 & -2.1851 & $13\pm4$ & 24.20 & NGC~5506 & 1.64 & 24.57 & 23.06 & 7.97 & 10.65 & 1.39 \\
    KiDS\_214.81\_3.12 & 214.8123 & 3.1243 & $23\pm5$ & 22.40 & NGC 5576 & 1.53 & 25.13 & 65.44 & 7.95 & 10.90 & 2.58 \\
    KiDS\_224.35\_2.33 & 224.3475 & 2.3256 & $7\pm3$ & 26.40 & Field & 1.35 & 25.59 & 36.32 & 7.76 & 10.38 & 1.22 \\
    KiDS\_226.08\_1.76$^{*}$ & 226.0799 & 1.7550 & $13\pm4$ & 26.40 & NGC 5846 & 1.64 & 24.74 & 42.25 & 7.77 & 10.65 & 2.21 \\
    KiDS\_226.33\_1.81$^{**}$ & 226.3342 & 1.8115 & $32\pm6$ & 24.55 & NGC 5846 & 2.17 & 25.70 & 60.77 & 8.01 & 11.04 & 3.13 \\
    KiDS\_346.30\_-29.81 & 346.2967 & -29.8066 & $9\pm3$ & 22.80 & Field & 1.61 & 25.18 & 21.73 & 7.88 & 10.54 & 1.19 \\
    KiDS\_347.06\_-28.18 & 347.0601 & -28.1830 & $9\pm3$ & 21.90 & NGC 7507 & 1.53 & 24.97 & 28.73 & 7.84 & 10.54 & 1.30 \\
    KiDS\_347.97\_-28.93 & 347.9699 & -28.9260 & $17\pm4$ & 21.90 & NGC 7507 & 1.57 & 25.23 & 78.54 & 8.33 & 10.79 & 0.80 \\
\hline
\end{tabular}%
}
\vspace{2pt}
\begin{minipage}{\textwidth}
\footnotesize
\textit{Notes.}
$N_{\rm GC}$ is the GC population inferred from
candidates selected within $2R_{\rm e}$ after correcting for GCLF
incompleteness. The halo masses $M_h$ are inferred from the GC
abundance using the empirical relation between GC system mass
and halo mass. The Environment column lists the host group or
cluster when both $R_{\rm proj}/r_{200}\leq1.5$ and
$|D-D_{\rm host}|\leq2~\mathrm{Mpc}$ are satisfied for that host.
Candidates with no matching host in the comparison catalogues
are listed as Field.\\
$^{*}$ Previously reported by Forbes et al. (2019) in NGC 5846.\\
$^{**}$ Previously reported by Marleau et al. (2021) in NGC 5846.\\
$^{***}$ Previously reported by Venhola et al. (2017) in Fornax.
\end{minipage}
\end{table*}

The two overlapping UDGs are FDS10\_LSB52 and FDS11\_LSB49. For FDS10\_LSB52, \citet{prole2019halo} reported a background-corrected GC population of $N_{\rm GC}=4.7^{+5.8}_{-3.4}$. For FDS11\_LSB49, they reported $N_{\rm GC}=4.4^{+5.8}_{-3.0}$. Applying our pipeline to the KiDS-DR5 images, we recover $N_{\rm GC}=6\pm3$ for FDS10\_LSB52 and $N_{\rm GC}=5\pm2$ for FDS11\_LSB49. In both cases, our measurements are consistent with the background-corrected estimates reported by \citet{prole2019halo}, and they agree particularly well with their quoted best estimates. This agreement provides an additional cross-check on the reliability of our GC-detection pipeline.

We note, however, that the finite depth of the KiDS imaging introduces an intrinsic limitation to our GC-detection pipeline. Owing to the relatively bright limiting magnitude of KiDS, a fraction of intrinsically faint globular clusters cannot be efficiently detected, resulting in an incomplete sampling of the faint end of the globular cluster luminosity function (GCLF). Consequently, the observed GC population is biased toward brighter magnitudes, and the directly detected GC counts represent a lower bound on the true GC population.

Since the GCLF is well described by a Gaussian distribution, we account for this incompleteness by extrapolating the observed GC counts toward the faint end following the expected Gaussian shape of the GCLF. Specifically, we assume a Gaussian GCLF with a dispersion of $\sigma = 1.0$~mag and apply a completeness correction to the observed GC counts accordingly \citep{miller2007globular,peng2016rich, lim2018globular}. This procedure allows us to estimate the contribution of undetected faint GCs and to mitigate the impact of the KiDS limiting magnitude on our measurements. The estimated contribution from the faint, undetected GC population is illustrated by the dashed component in the upper-right panel of Fig.~\ref{fig:gc_validation_ngc5846}.

In addition, we tested the robustness of our distance estimates by extending the GC detection region from $2R_{\rm e}$ to $3R_{\rm e}$ and constructing the corresponding GCLFs using the same pipeline. This enlarged aperture allows us to include a larger number of GC candidates and to assess whether the inferred GCLF properties depend sensitively on the adopted search radius. Among the 24 UDGs examined, 15 show an increase in the number of detected GC candidates, while the remaining nine show no change. Overall, the total number of detected GC candidates increases by approximately 10\% when extending the search radius from $2R_{\rm e}$ to $3R_{\rm e}$. We find that, for the majority of UDGs, the distances derived from the $3R_{\rm e}$ GCLFs are in good agreement with those obtained from the $2R_{\rm e}$ analysis. For the small number of systems showing differences, the discrepancies remain within 5\%. No systematic offset is observed between the two sets of distance estimates. These results indicate that, although increasing the search radius leads to a modest increase in the detected GC population, the GCLF-based distance estimates remain relatively insensitive to the adopted search radius, supporting the robustness of our distance measurements within $2R_{\rm e}$.

\subsection{Application to the HQ KiDS UDG catalog}
\label{sec:Application to the HQ KiDS UDG catalog}

The validated GC-detection pipeline is applied to the full high-quality (HQ) UDG catalog previously identified in KiDS. For each UDG candidate, we search for associated GC candidates and, where possible, construct the globular cluster luminosity function (GCLF) to derive distance estimates. Since a reliable characterization of the GCLF requires a minimum number of detected GC candidates, we adopt a minimum threshold of three detected candidates, prior to completeness correction, for inclusion in the subsequent analysis. This threshold serves as a practical sample-selection criterion rather than a definition of GC richness. We emphasize that UDG candidates with fewer than three GC detections are not excluded from the HQ catalog and remain bona fide UDG candidates; however, they are not considered further in the present analysis owing to the lack of sufficient GC statistics. Applying this selection, we obtain a subsample of 55 UDG candidates for which GCLF-based distance estimates can be attempted. Among these, 24 systems have physical effective radii larger than 1.3~kpc and constitute our final sample.

\begin{figure*}[ht!]
    \centering
    \includegraphics[width=0.95\textwidth]{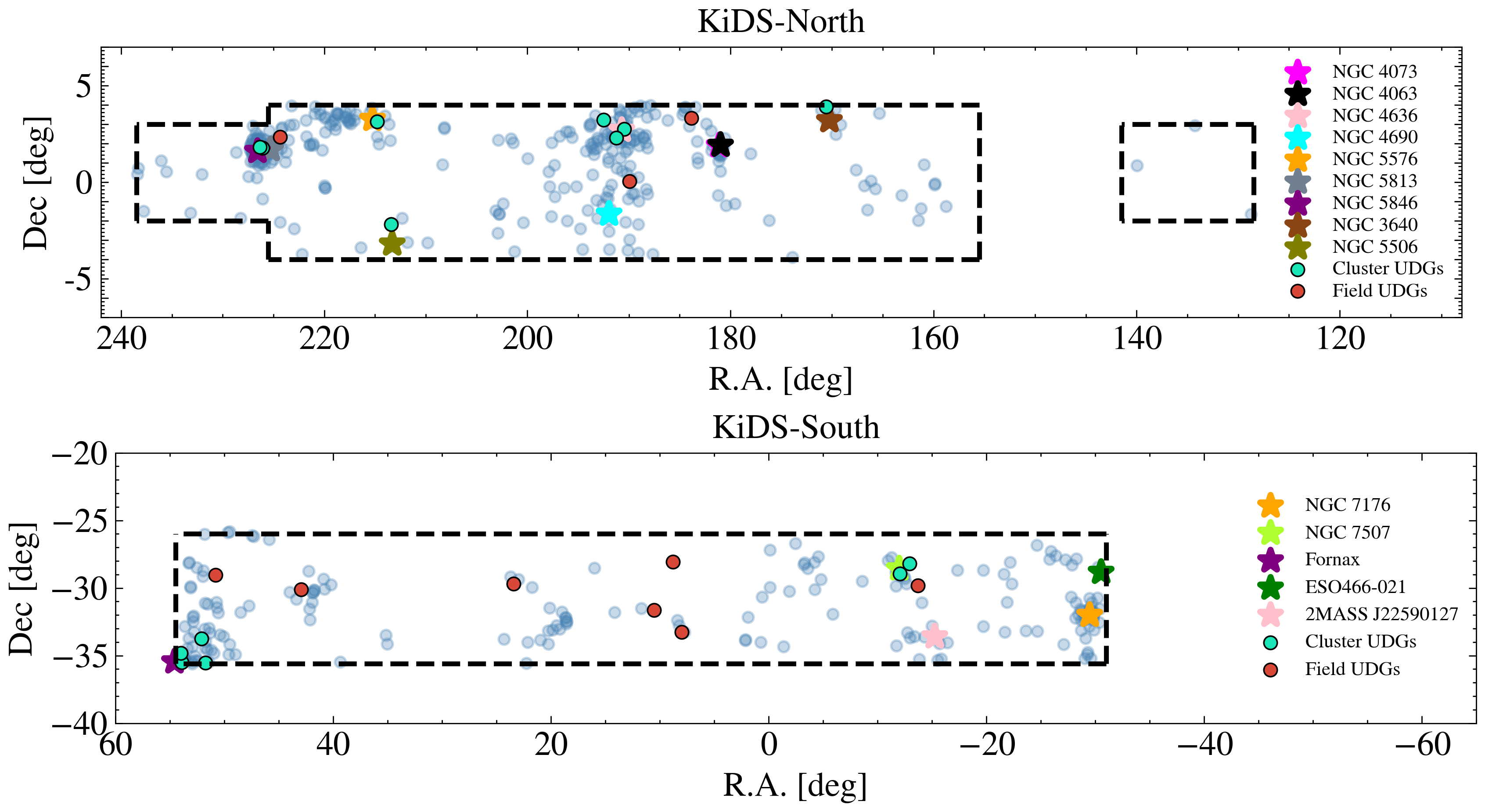}
    \caption{Sky distribution of the 24 GC-rich UDG candidates studied here in KiDS-North (top) and KiDS-South (bottom). Red circles mark the 10 field candidates, and turquoise circles with mark the 14 candidates associated with galaxy groups or clusters. Faint blue circles show the high-quality (HQ) UDG
    candidates from our previous KiDS study for spatial context. The full HQ catalog is presented in \citet{su2026ultra}.}
    \label{fig:distribution_24_udg}
\end{figure*}

Recent studies commonly classify UDGs as GC-rich when their total GC populations reach approximately $N_{\rm GC}\gtrsim20$ \citep{forbes2025some,buzzo2025multiple}. However, the limited photometric depth and seeing conditions of the KiDS survey restrict the detection of faint GCs, and our pipeline may therefore recover predominantly the bright end of the underlying GCLF. As a result, the number of detected GC candidates may substantially underestimate the intrinsic GC population. Considering these observational limitations, we adopt a more inclusive observational criterion of $N_{\rm GC,det}\gtrsim10$ to identify potentially GC-rich systems. Among our final sample of 24 UDG candidates, 15 systems satisfy this criterion and are therefore identified as potentially GC-rich UDG candidates. We emphasize that this classification is tentative, as reliable estimates of their total GC populations cannot be obtained from the current data owing to the limited detection depth and the uncertainties associated with completeness corrections, background contamination, and extrapolation of the observed GCLFs. Consequently, these 15 systems should be regarded as promising candidates that may satisfy the conventional GC-rich criterion, rather than confirmed GC-rich UDGs.

In this work, we retain UDG candidates with physical effective radii $R_{\rm e}>1.3~\rm kpc$. This slightly more inclusive threshold is adopted for consistency with our parent KiDS UDG catalog and to account for small systematic uncertainties in the growth-curve measurements. Furthermore, recent studies suggest that galaxies slightly below the conventional UDG size threshold of $R_{\rm e}=1.5~\rm kpc$ may exhibit physical properties similar to those of classical UDGs. In particular, \citealt{buzzo2025multiple} investigated a sample of 36 nearly ultra-diffuse galaxies (NUDGes) and found that, despite their generally smaller sizes and higher surface brightnesses, these systems share broadly similar stellar population properties and trends in their globular cluster populations with UDGs. These findings suggest that the conventional size threshold does not necessarily represent a sharp physical distinction between UDGs and slightly smaller low-surface-brightness galaxies, providing additional motivation for adopting a more inclusive size criterion in our analysis.

As discussed in our original catalog paper \citep{su2026ultra}, the growth-curve method introduces a small systematic underestimate in the derived galaxy luminosities, with an offset of $\sim0.08~mag$. If this offset is interpreted as an offset in distance modulus, the inferred distance scales as $D'/D = 10^{0.2\Delta\mu}$. For $\Delta\mu=0.08~mag$, this gives $D'/D \simeq 10^{0.016} \simeq 1.038$, corresponding to an increase of about 3.8\% in distance. Since the physical effective radius scales as $R_{\rm e}=\theta D$, the inferred radius would increase by the same factor, $R'_{\rm e}/R_{\rm e}\simeq1.038$. Therefore, the original physical radii are expected to be slightly underestimated in a systematic sense. Although this correction alone does not move all borderline systems above the canonical 1.5~kpc threshold, it supports our decision to keep candidates close to the boundary rather than applying a sharp cut. For consistency, the effective radii shown in Fig.~\ref{fig:parameter_distribution} are the corrected values, obtained by applying this (1.038) scaling factor to the original estimates.

The distributions of key structural and photometric parameters for the final sample of 24 UDG candidates are shown in Fig.~\ref{fig:parameter_distribution}, demonstrating that these systems occupy the same region of parameter space as UDGs identified in previous surveys. In particular, our sample overlaps well with the MATLAS UDG population \citep{marleau2021ultra, marleau2024dwarf}, whose 74 UDGs span a wide range of environments from low-density field regions to galaxy groups and, in some cases, higher-density environments. The inclusion of candidates with effective radii slightly below the conventional 1.5~kpc threshold also allows us to investigate UDG-like systems near the traditional selection boundary, consistent with the similarities between UDGs and NUDGes reported by \citealt{buzzo2025multiple}. A particularly important aspect of our sample is that it contains a statistically relevant subset of field UDG candidates. This is a key point of the present work, since field UDG samples remain more limited than their cluster counterparts in the literature, and our sample therefore provides a useful basis for a statistical investigation of diffuse galaxies in low-density environments. At the same time, the full sample still includes systems associated with groups and denser environments, enabling environmental trends to be explored across a broad range of conditions. The main properties of these 24 candidates are summarized in Table~\ref{table1}. In the following sections, we use this refined sample to compare the structural and photometric properties of UDGs in field and cluster environments.

\section{Analysis}
\label{analysis}
\subsection{Spatial distribution and environmental association}
We examine the spatial distribution of the 24 UDG candidates across
the KiDS footprint. Their projected positions on the sky are shown
in Fig.~\ref{fig:distribution_24_udg}. To assess
their environments, we cross-match the UDG candidates with nearby
galaxy groups and clusters. For each candidate--host pair, we
calculate the projected separation $R_{\rm proj}$ from the host
center using the angular separation and the adopted host distance.
We then normalize $R_{\rm proj}$ by the host radius $R_{200}$,
defined as the radius within which the mean enclosed density is
200 times the critical density of the Universe.

For NGC~4636 and NGC~5846, we derive $R_{200}$ from
the X-ray-based $M_{200}$ estimates of \citet{bohringer2021cosmic},
using $R_{200}=[3M_{200}/(4\pi\,200\rho_{\rm crit})]^{1/3}$.
For the other comparison hosts, including Fornax, we estimate
$R_{200}$ from the projected second-turnaround radii $R_{2t}$
tabulated by \citet{kourkchi2017galaxy}. Using the scaling relations
of \citet{tully2015galaxy} evaluated at the same halo mass, we adopt
$R_{200}\simeq(0.198/0.215)R_{2t}\simeq0.92R_{2t}$.

We classify a UDG as group- or cluster-associated
if it satisfies both $R_{\rm proj}/R_{200}\leq1.5$ and
$|D_{\rm GCLF}-D_{\rm host}|\leq2~\mathrm{Mpc}$ for the same host.
The projected search radius of $1.5R_{200}$ allows us to include
systems in the outskirts of groups and clusters. This choice is
motivated in part by the presence of backsplash galaxies, which can
move beyond the virial radius after passing through their host halo
\citep{gill2004evolution,bahe2013does,benavides2021quiescent}.
The distance criterion helps reduce chance associations along the
line of sight. Candidates for which no host in the comparison
catalogs satisfies both criteria are classified as
\emph{field UDGs}.

These criteria yield 14 group- or cluster-associated
UDGs and 10 field UDGs. Of the 14 associated candidates, nine have
$R_{\rm proj}\leq R_{200}$ and five lie at projected separations
of $R_{200}<R_{\rm proj}\leq1.5R_{200}$. The orbital histories
of these five outer systems remain uncertain given the available
projected positions and distance estimates. For brevity, we refer
to the combined group- and cluster-associated subsample as
\emph{cluster UDGs} hereafter.

Four of the 14 cluster UDGs were previously reported
in the literature \citep{venhola2017fornax,marleau2021ultra}, as
indicated in Table~\ref{table1}. For eight of the 14 systems, the
host associations found here agree with those identified in our
earlier KiDS analysis \citep{su2026ultra}. Our GC measurements
also place tighter constraints on the GC populations of two
previously reported systems, FDS10\_LSB52 and FDS11\_LSB49.

To complement the host-based environmental
classification, we estimate the local galaxy number density
following the nearest-neighbor approach described in Appendix~A
of \citealt{buzzo2025multiple}. We select neighboring galaxies from the
2MASS Redshift Survey (2MRS; \citealt{huchra20122mass}) with
extinction-corrected apparent magnitudes $K_s<10.5$ and valid
positive radial velocities. For each UDG, we use its sky position
and GCLF-based distance, while neighbor distances are approximated
as $D=v/H_0$, adopting
$H_0=70~\mathrm{km\,s^{-1}\,Mpc^{-1}}$. We calculate the
three-dimensional separations and define
$\rho_{10}=10/[(4/3)\pi r_{10}^{3}]$, where $r_{10}$ is the
distance to the tenth nearest neighbor. This estimator provides
a continuous measure of environment for comparison with
$N_{\rm GC}$. Our implementation follows the neighbor selection
and density definition of \citealt{buzzo2025multiple}, with the distance
treatment specified above. The resulting estimates are sensitive
to peculiar velocities, catalog completeness, and uncertainties
in the GCLF-based distances. These effects are not propagated
into formal density uncertainties here.

Writing $\eta_{10}\equiv\log_{10}[\rho_{10}/(\mathrm{Mpc}^{-3})]$,
we obtain a median of $-1.07$ for the 24 UDG candidates.
The mean $\eta_{10}$ is $-0.98$ for the group- or
cluster-associated candidates and $-1.51$ for the field
candidates. For a descriptive comparison, we adopt
$\eta_{10}=-1.0$, approximately the sample median, as an
operational boundary between relatively higher- and lower-density
environments. This division yields ten candidates with
$\eta_{10}\geq-1.0$ and 14 with $\eta_{10}<-1.0$.
The higher-density subset contains eight group- or
cluster-associated candidates and two field candidates,
while the lower-density subset contains six associated
candidates and eight field candidates. We use this density
measure as a complementary environmental diagnostic and retain
the cluster and field classifications defined above for the
subsequent analyses.

\subsection{GC system and Specific frequency}
\begin{figure}[htbp]
    \centering
    \includegraphics[width=0.45\textwidth]{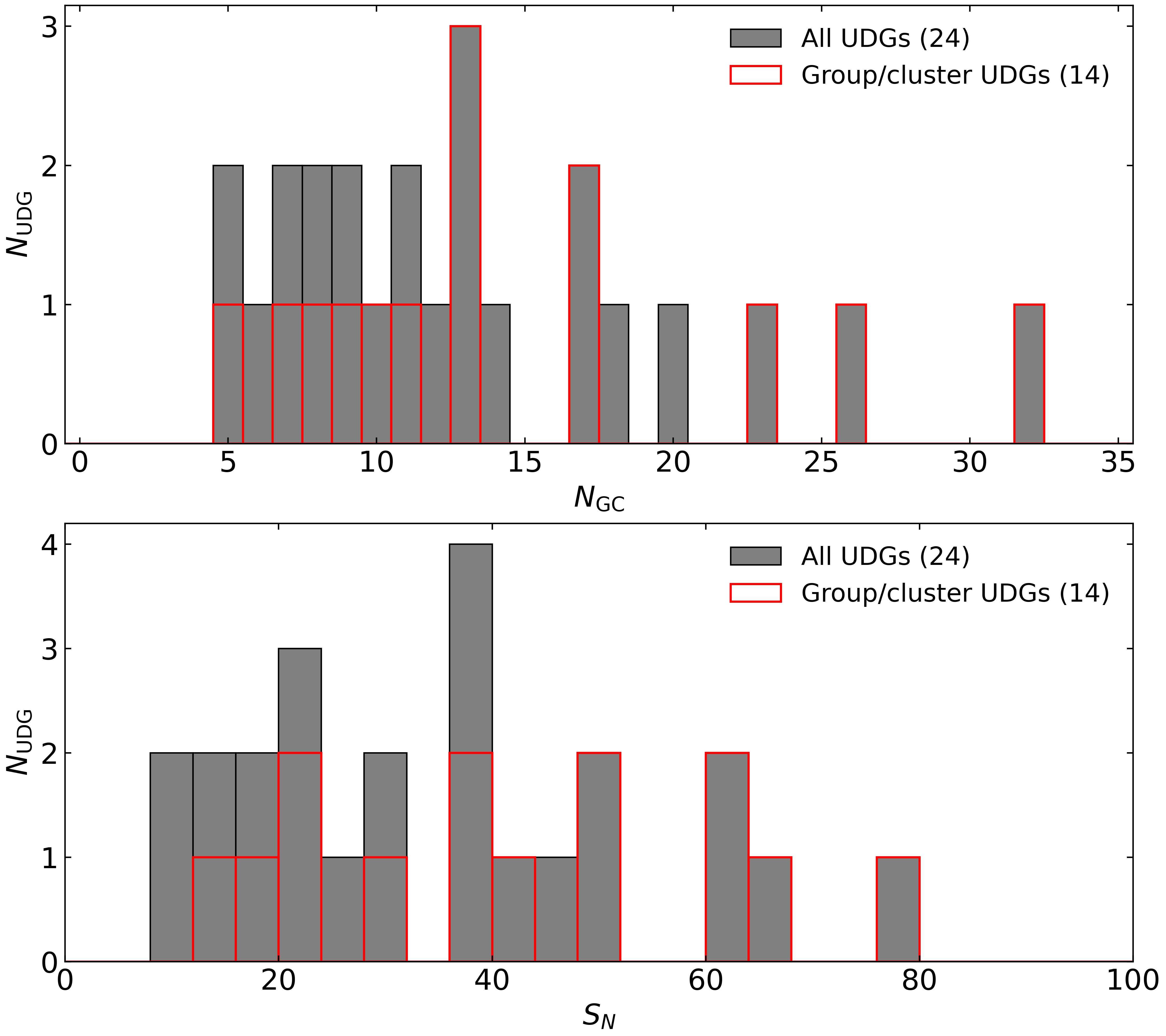}
    \caption{Environmental trends of the richness of globular cluster (GC) systems in UDGs.
        \emph{Top panel:} Distribution of the completeness-corrected number of globular clusters, $N_{\rm GC}$, for the full UDG sample, with the contribution from cluster-associated UDGs highlighted.
        \emph{Bottom panel:} Distribution of the GC specific frequency, $S_N$, for the same sample.}
    \label{fig:Sn_Ngc}
\end{figure}

We next investigate the properties of the globular cluster (GC)
systems associated with the 24 UDG candidates, focusing on the
distribution of GC numbers and the corresponding specific frequencies.
The distribution of the GCLF-completeness-corrected
GC counts, $N_{\rm GC}$, is shown in Fig.~\ref{fig:Sn_Ngc}.
The 14 UDGs associated with groups or clusters have
a mean GC count of $\langle N_{\rm GC}\rangle_{\rm cluster}=14.57$,
compared with $\langle N_{\rm GC}\rangle_{\rm field}=11.00$ for the
10 field UDGs. Five of the 14 group- or cluster-associated UDGs
(36\%) and two of the 10 field UDGs (20\%) have $N_{\rm GC}>15$.
The estimated GC counts overlap substantially between the two
subsamples, so we interpret these differences as descriptive trends
in the present sample.

\begin{figure}
    \centering
    \includegraphics[width=0.45\textwidth]{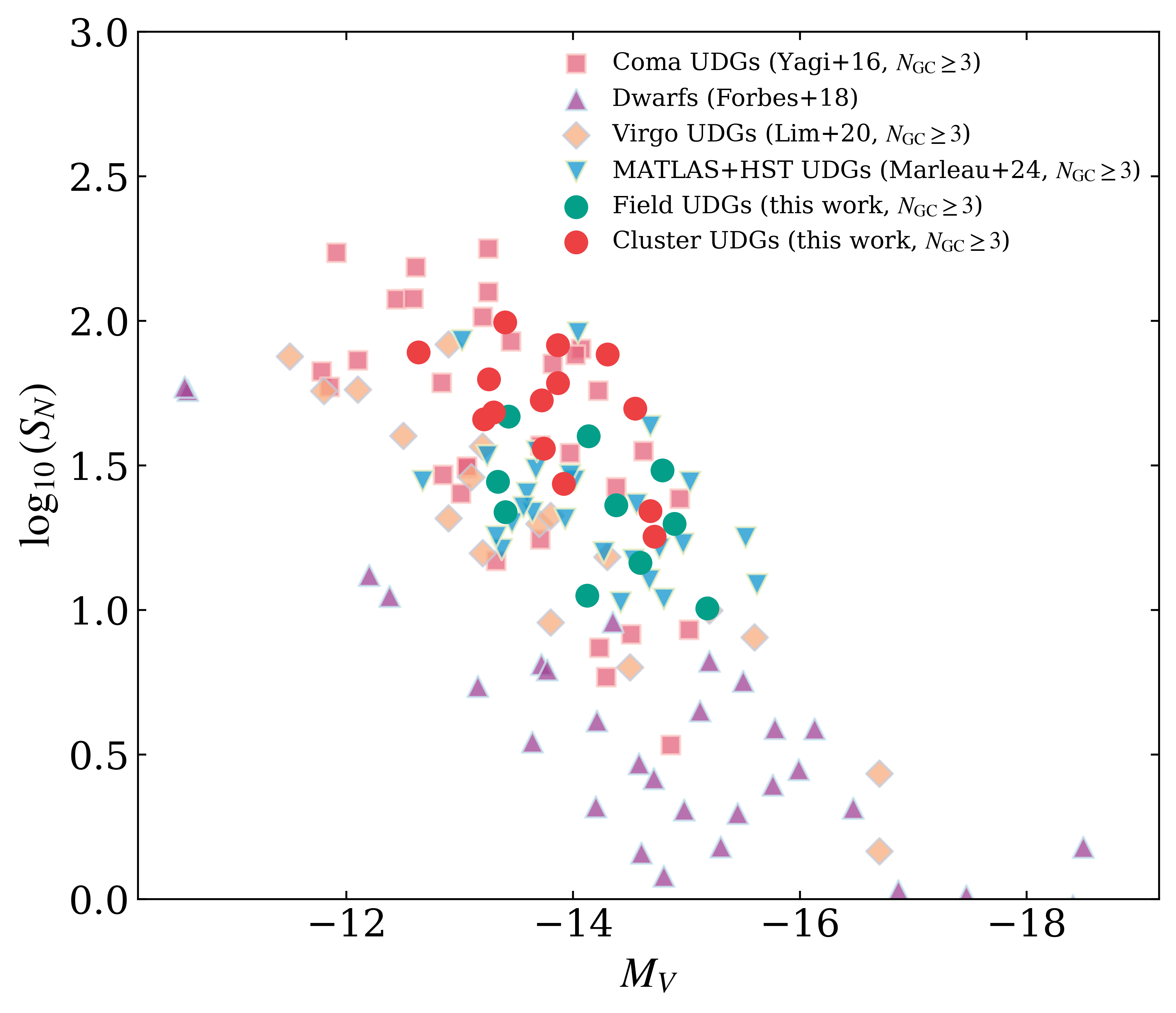}
    \caption{Globular cluster specific frequency as a function of host-galaxy luminosity. The figure shows $\log_{10} S_N$ versus absolute $V$-band magnitude, $M_V$, for UDGs identified in this work, separated into cluster and field systems, together with comparison samples of UDGs in the Coma \citep{lim2018globular} and Virgo \citep{lim2020next} clusters, as well as UDGs from the MATLAS+HST survey \citep{marleau2024dwarf} and dwarf galaxies from \citep{forbes2018extending}. 
    }
    \label{fig:Mv-Sn}
\end{figure}

This pattern suggests that environment may influence the subsequent evolution of UDGs, but not necessarily by increasing their GC populations after infall. A more plausible interpretation is that GC richness is linked primarily to early halo assembly, while environmental processes such as quenching, preprocessing, tidal stripping, and differential survival shape the present-day properties of the galaxies. In this picture, systems that undergo stronger environmental processing may lose part of their GC systems, with the stripped clusters potentially contributing to the intragroup or intracluster GC population, or be transformed to the point that they no longer appear as UDGs, whereas the cluster UDGs that remain observable today may preferentially be those that have preserved both their diffuse stellar structure and a significant fraction of their original GC population.

To further quantify the richness of the GC systems relative to the stellar content of the host galaxies, we compute the GC specific frequency, $S_N$, defined as
\begin{equation}
S_N = N_{\rm GC} \times 10^{0.4 (M_V + 15)},
\end{equation}
where $N_{\rm GC}$ is the total number of globular clusters associated with a given UDG, and $M_V$ is the absolute magnitude of the galaxy in the $V$ band \citep{harris1981globular}.

The $V$-band absolute magnitudes are derived from the observed photometry using a color-dependent transformation based on the $g-r$ color. Specifically, we adopt
\begin{equation}
V = g - 0.59 (g - r) - 0.01,
\end{equation}
following standard empirical color transformations appropriate for old stellar populations. Absolute magnitudes are then obtained using the distances inferred from the GCLF analysis.

The resulting distribution of $S_N$ is shown in Fig.~\ref{fig:Sn_Ngc}.
As with $N_{\rm GC}$, higher $S_N$ values are more
common among the UDGs associated with galaxy groups or clusters
in our sample. Seven of the 14 group- or cluster-associated
candidates (50\%) have $S_N>40$, compared with one of the 10
field candidates (10\%). The median $S_N$ is 40.86 for the
group- or cluster-associated candidates and 24.78 for the field
candidates. This pattern is broadly consistent with previous
studies of GC-rich UDGs in nearby clusters
(e.g. \citealt{amorisco2018globular,lim2018globular,marleau2021ultra}).
The $S_N$ distributions overlap substantially, so we treat the
difference as a descriptive trend in the present sample.

We further find that the median value of the cluster subsample, $S_N = 40.86$, is higher than the median value of $S_N = 10.7$ reported for UDGs in the Perseus cluster \citep{marleau2025euclid}, while remaining comparable to the median values measured for UDG populations in the Virgo \citep{lim2020next} and Coma clusters \citep{lim2018globular} ($S_N = 42.4$ and $S_N = 30.6$, respectively). The $S_N$ values of our cluster UDGs span a range of $14.27$--$78.54$, remaining well below the most extreme specific frequencies measured for Virgo and Coma UDGs ($S_N = 187.1$ and $S_N = 348.3$, respectively).

\subsection{$S_{N}$ versus host-galaxy luminosity}
\label{sn_mv}
To further explore the dependence of globular cluster (GC) richness on host-galaxy properties, we examine the relation between GC specific frequency and galaxy luminosity. In Fig.~\ref{fig:Mv-Sn}, we plot $\log S_N$ as a function of the absolute $V$-band magnitude, $M_V$, for our UDG sample. The figure includes both cluster and field UDGs identified in this work, together with comparison samples of UDGs in the Coma Cluster from \citet{yagi2016catalog}, for which the GC numbers were taken from \citet{lim2018globular}, as well as UDGs in the Virgo Cluster \citep{lim2020next}, MATLAS+HST UDGs \citep{marleau2024dwarf}, and dwarf galaxies spanning a similar luminosity range \citep{forbes2018extending}.

Our cluster UDGs occupy a region of the $S_N$--$M_V$ plane that is broadly consistent with that defined by UDGs in the Coma and Virgo clusters. Field UDGs show a tendency toward lower GC specific frequencies at fixed luminosity, suggesting a systematic offset with respect to the cluster population. However, they still display a substantial overlap with the Virgo and Coma UDGs, rather than lying clearly along the locus of normal dwarf galaxies. At the same time, the Virgo and Coma samples themselves include a tail of objects with relatively low $S_N$ values, comparable to those of normal dwarfs and to some of the field UDGs in our sample. This suggests that GC richness in UDGs is not determined solely by present-day environment, but may also depend on the timing and degree of environmental processing. In this context, cluster UDGs with low $S_N$ may represent systems located in the outskirts of clusters or relatively recent arrivals that still retain more field-like GC properties.

At fixed host-galaxy luminosity, UDGs as a class exhibit higher GC specific frequencies than typical dwarf galaxies (e.g. \citealt{peng2008acs, harris2017galactic}). This offset indicates that, relative to their stellar content, UDGs host disproportionately rich GC systems. A plausible interpretation is that UDGs experienced highly efficient GC formation at early epochs, followed by suppressed or prematurely quenched field-star formation \citep{beasley2002formation, kruijssen2014globular}. As a consequence, while the stellar mass growth of the host galaxy was curtailed, the initially formed GC population was largely preserved, leading to elevated $S_N$ values.

A well-known trend of increasing $S_N$ with decreasing luminosity among low-mass galaxies is clearly visible in Fig.~\ref{fig:Mv-Sn}. At higher galaxy masses than those probed here, this relation is known to reverse, with $S_N$ increasing again toward higher luminosities, giving rise to the characteristic U-shaped distribution reported in previous studies \citep{forbes2005bimodal, georgiev2010globular}. The UDGs in our sample follow the general trend observed for dwarf galaxies; however, similarly to low-surface-brightness (LSB) dwarfs, they reach significantly higher specific frequencies at a given host-galaxy luminosity \citep{van2016abundance}. Nevertheless, the substantial scatter in $S_N$ at fixed luminosity suggests that environment does not act in a uniform way across all UDGs. Instead, the observed distribution may reflect a combination of early GC formation efficiency, quenching history, infall time, and the degree of environmental processing. This behavior suggests that UDGs do not constitute a distinct class in terms of their GC scaling relations, but rather represent the high-$S_N$ extension of the dwarf galaxy population, particularly in dense environments where early formation and environmental processing are most effective, while systems with lower $S_N$ may trace more weakly processed or more recently accreted populations.

\subsection{GC population versus stellar mass}
\label{sn_stellarmass}
Before examining the dependence of globular cluster (GC) richness on stellar mass, we first describe how the stellar masses of the UDGs in our sample are estimated. Stellar masses are derived from optical photometry following a standard luminosity--mass conversion. We compute the absolute magnitude in the $r$ band as

\begin{equation}
M_r = m_r - 5 \log_{10}\!\left(\frac{D}{10\,{\rm pc}}\right),
\end{equation}

where $m_r$ is the extinction-corrected apparent $r$-band magnitude of the UDG and $D$ is the distance to the galaxy. Distances are primarily taken from the GCLF analysis when available. The corresponding $r$-band luminosity is then obtained through

\begin{equation}
\frac{L_r}{L_{r,\odot}} = 10^{-0.4\,(M_r - M_{r,\odot})},
\end{equation}

where $M_{r,\odot}$ is the absolute $r$-band magnitude of the Sun.

We estimate the $r$-band stellar mass-to-light ratio
of each UDG from its extinction-corrected color $(g-r)_0$ using
the relation of \citet{bell2003optical}.
\begin{equation}
\log_{10}\!\left(\frac{M_\star/L_r}{M_\odot/L_{r,\odot}}\right)
= -0.306 + 1.097\,(g-r)_0 .
\end{equation}
The adopted mass-to-light ratios range from $1.12$
to $6.50\,M_\odot/L_{r,\odot}$, with a median of
$1.95\,M_\odot/L_{r,\odot}$. We obtain the initial stellar-mass
estimate by multiplying each UDG's $r$-band luminosity by its
corresponding mass-to-light ratio.
\begin{equation}
M_\star = (M_\star/L_r)\,L_r .
\end{equation}

\begin{figure}[]
    \centering
    \includegraphics[width=0.45\textwidth]{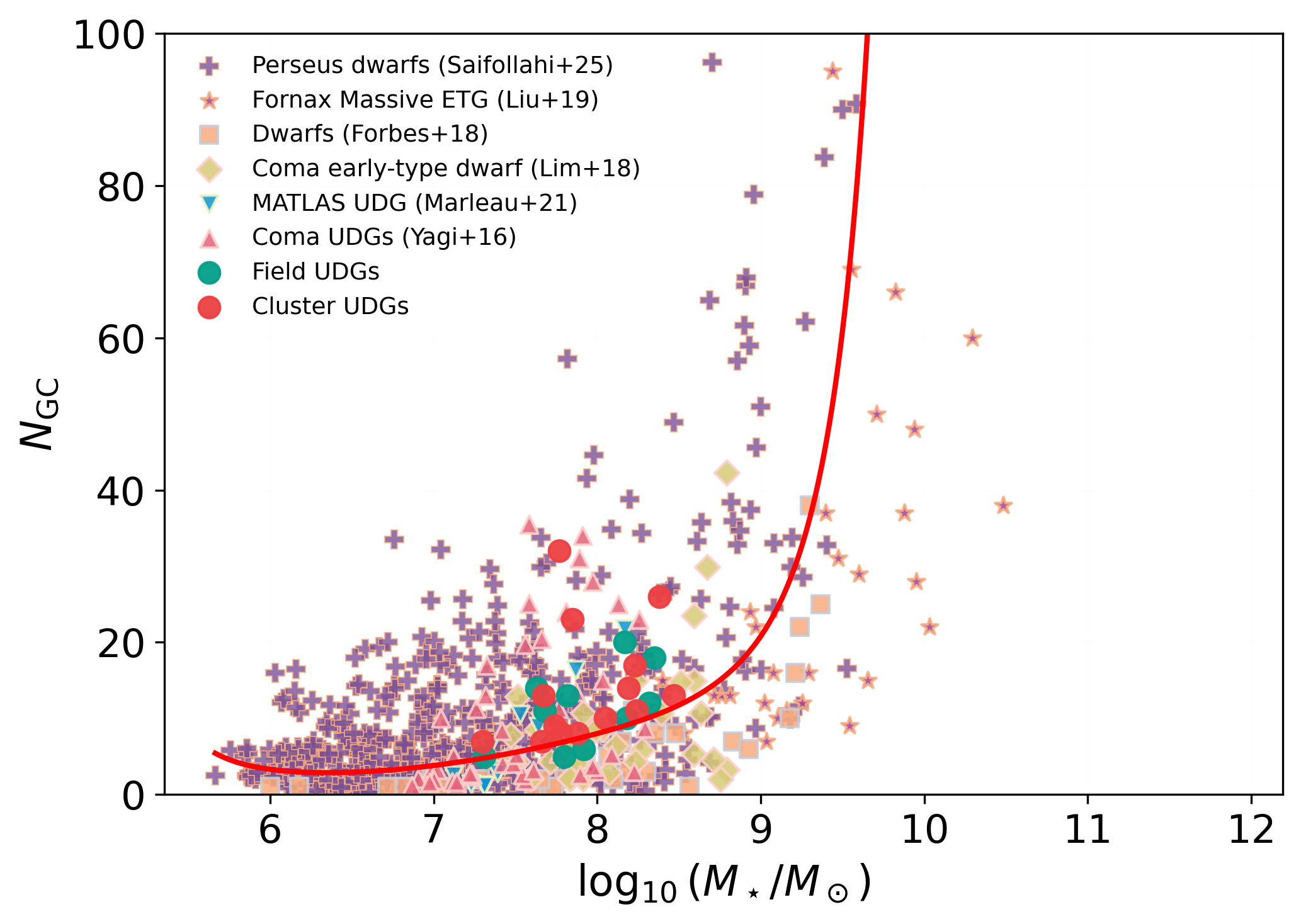}
    \caption{Total number of globular clusters, $N_{\rm GC}$, as a function of stellar mass, $M_\star$, for the UDG sample analyzed in this work. Field and cluster UDGs are shown separately and compared with literature samples, including Coma early-type dwarf galaxies \citep{lim2020next}, Coma UDGs from \citep{yagi2016catalog}, dwarf galaxies from the SLUGGS survey \citep{forbes2018extending}, massive early-type galaxies from the HST/ACS Fornax Cluster Survey \citep{liu2019acs}, and UDGs from the MATLAS survey \citep{marleau2021ultra}. The red solid curves represent fourth-order polynomial fits to the $N_{\rm GC}$--$M_\star$ relation defined by the UDGs shown in the figure.
    }
    \label{fig:Ngc_stellarmass}
\end{figure}

At fixed stellar mass, cluster UDGs tend, on average, to populate the upper envelope of the $N_{\rm GC}$--$M_\star$ relation, whereas field UDGs more often lie closer to its lower part and to the locus traced by typical dwarf galaxies. As shown in Fig.~7, our UDGs broadly overlap with literature samples of dwarf galaxies and UDGs, while exhibiting considerable scatter in GC richness at a given stellar mass. In particular, UDGs with unusually large GC populations for their stellar mass are more frequently associated with groups and clusters, suggesting that dense environments preferentially host or preserve the most GC-rich UDGs. At the same time, the overlap between field and cluster systems remains substantial, and the present sample does not support a clean statistical separation between the two populations. This behavior broadly mirrors the trends observed in the $S_N$--$M_V$ plane (Section~\ref{sn_mv}), indicating that any environmental dependence is present but relatively modest.

This behavior is consistent with previous studies of UDGs in the MATLAS survey \citep{marleau2021ultra}, where GC-rich systems were found to populate the upper envelope of the dwarf-galaxy $N_{\rm GC}$--$M_\star$ relation. Such an offset can be naturally explained if UDGs experienced efficient GC formation at early epochs, followed by suppressed or prematurely quenched field-star formation. In this scenario, the stellar mass growth of the host galaxy is curtailed, while the initially formed GC population is largely preserved, leading to elevated GC numbers relative to stellar mass. Our results therefore reinforce the view that UDGs are not a distinct class in terms of their GC scaling relations, but rather represent the high-$N_{\rm GC}$ extension of the dwarf-galaxy population.

The dependence of $N_{\rm GC}$ on stellar mass, together with the large scatter observed at fixed $M_\star$, suggests that stellar mass alone may not be the fundamental parameter governing the GC content of UDGs. Instead, these trends naturally point toward a closer connection between GC populations and the underlying dark matter halo mass, $M_{\rm h}$. In particular, the elevated GC numbers observed in cluster UDGs at a given stellar mass imply that these systems may reside in disproportionately massive halos compared to their field counterparts.

This interpretation is supported by a growing body of observational and theoretical work indicating that the total number or total mass of GCs correlates more tightly with halo mass than with stellar mass (e.g. \citealt{blakeslee1999globular,spitler2009new,forbes2020globular}). In this framework, GC populations are thought to trace the early assembly of dark matter halos, while the stellar component is more strongly affected by subsequent environmental processes such as gas stripping, feedback, and quenching.

Motivated by these considerations, in the following section we investigate the connection between GC richness and halo mass for our UDG sample, and explore the implied GC formation efficiency across different environments.

\subsection{Stellar mass versus halo mass}
Near-linear empirical relations have been established between the total halo mass of a galaxy and both the number of globular clusters and the total mass of the globular cluster system in normal galaxies \citep{spitler2009new, harris2017galactic, burkert2020high}. One of the best-studied examples among Coma UDGs is DF44. As discussed by \citet{burkert2020high}, the halo mass inferred from the radially extended stellar kinematics of DF44 provides an important comparison with GC-based estimates, although the level of agreement depends on the adopted GC abundance: the revised lower value of $N_{\rm GC}$ from \citet{saifollahi2021number} yields better consistency than the larger GC population originally reported by \citet{van2017abundance}.


In this work, we convert the observed GC number counts into estimates of the total halo masses of UDGs using an alternative empirical scaling relation between globular cluster systems and dark matter haloes. Following \citet{harris2017galactic}, the virial halo mass, $M_{\rm h}$ of a galaxy is related to the total mass of its globular cluster system, $M_{\rm GC,tot}$, through the relation
\begin{equation}
\frac{M_{\rm GC,tot}}{M_{\rm h}} = 2.9 \times 10^{-5}.
\end{equation}

To estimate $M_{\rm GC,tot}$, we adopt the approach of \citet{marleau2021ultra} and assume a mean globular cluster mass of $\langle M_{\rm GC} \rangle = 1 \times 10^{5}\,M_\odot$, which is appropriate for dwarf galaxies \citep{harris2017galactic}. The total GC system mass is then computed as 
\begin{equation}
M_{\rm GC,tot} = N_{\rm GC} \times \langle M_{\rm GC} \rangle .
\end{equation}

Combining these relations, the virial halo mass can be expressed as
\begin{equation}
M_{\rm h} = 3.45 \times 10^{9} \, N_{\rm GC} \, M_\odot .
\end{equation}

We also estimate the ratio of GC system mass to
stellar mass, $M_{\rm GCS}/M_\star$, where
$M_{\rm GCS}\equiv M_{\rm GC,tot}$. Adopting a mean GC mass of
$10^5\,M_\odot$, we obtain ratios ranging from 0.35\% to 3.13\%,
with a median of 0.92\% across the 24 UDG candidates. The median
is 1.09\% for the 14 group- or cluster-associated candidates and
0.77\% for the 10 field candidates, a difference of 0.32 percentage
points.

These ratios are broadly comparable to those
reported for other UDGs and low-surface-brightness galaxies.
The median halo-to-stellar mass ratio reported by
\citet{marleau2024dwarf} corresponds to a GC system-to-stellar
mass ratio of approximately 1.37\% when converted using the same
relation between GC system mass and halo mass.
\citet{janssens2024piper} found GC mass fractions of around 0.4\%
for low-surface-brightness galaxies in the Perseus cluster, with
higher values for some UDGs. \citet{janssens2022globular} reported
a fraction of approximately 2\% for the isolated UDG DGSAT~I.
Differences in sample selection, GC completeness, and stellar
mass estimates affect direct comparisons between these samples.

The resulting halo-mass distributions for cluster and field UDGs are shown in Fig.~8. A mild environmental trend is present, with cluster UDGs somewhat more frequently populating the high-mass end of the halo-mass distribution, whereas field UDGs extend to slightly lower halo masses. The mean halo mass of cluster UDGs is $\langle \log_{10}(M_h/M_\odot)\rangle = 10.63$, compared to $\langle \log_{10}(M_h/M_\odot)\rangle = 10.57$ for field UDGs. This difference is small, however, and the overlap between the two populations remains substantial.

This behavior is broadly consistent with the trends observed in $S_N$ and $N_{\rm GC}$ at fixed stellar mass (Sections~4.3 and 4.4), and suggests that any environmental dependence in the inferred halo masses is at most modest in the present sample. Rather than indicating a sharp dichotomy between field and cluster UDGs, the observed distribution may reflect a combination of early halo assembly, quenching history, preprocessing, infall time, and survival in dense environments. In this picture, UDGs that entered dense environments earlier may have experienced more prolonged environmental processing, whereas more recently accreted systems may still retain properties similar to those of field UDGs.

\begin{figure}[htbp]
    \centering
    \includegraphics[width=0.45\textwidth]{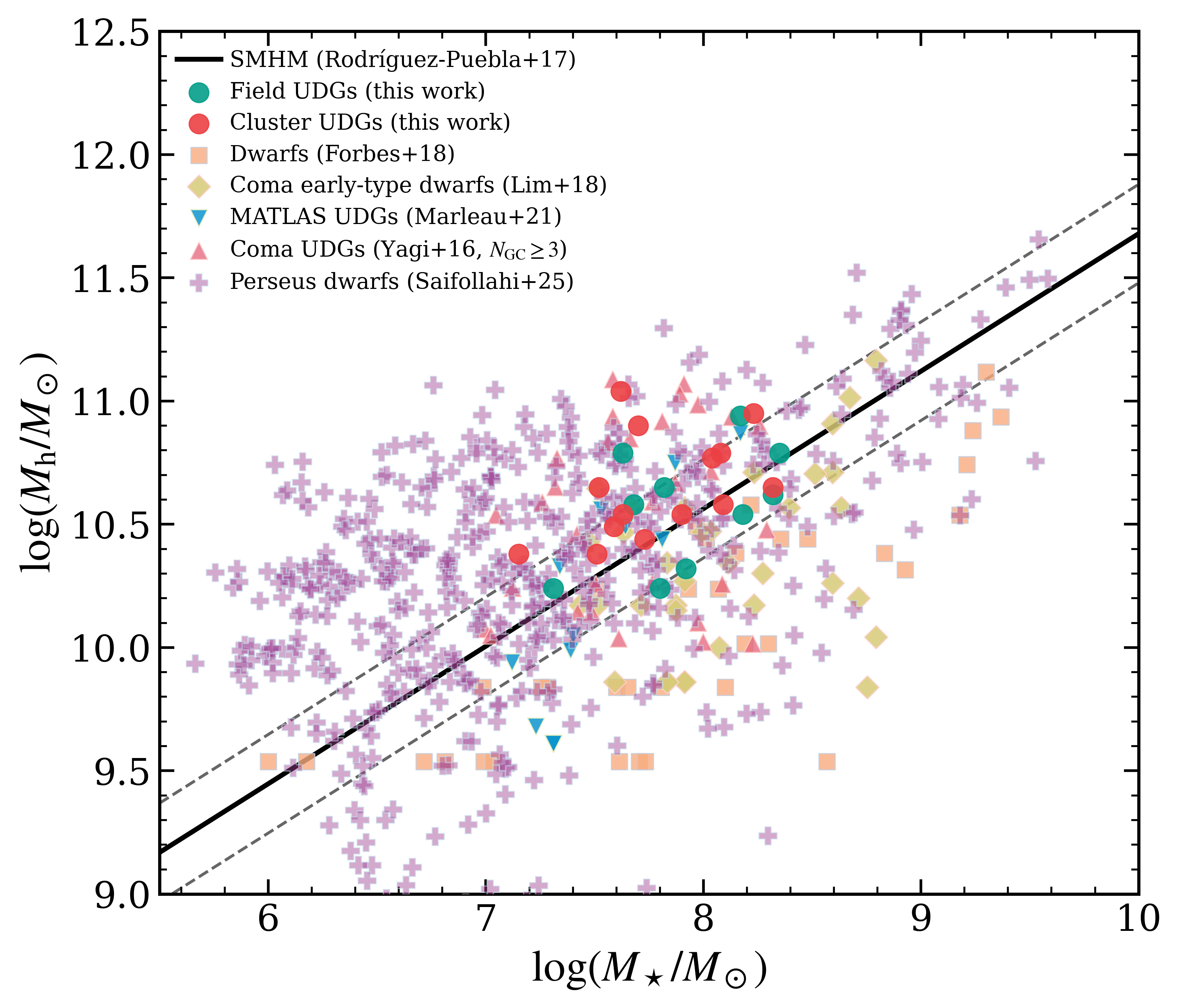}
    \caption{Halo mass versus stellar mass for the UDG sample analysed in this work. The halo mass is derived from the empirical scaling between GC system number and the total halo mass. The solid line shows the extrapolated stellar mass-halo mass (SMHM) relation for normal galaxies from \citet{rodriguez2017constraining}. The dashed lines indicate the typical intrinsic scatter of $\pm 0.2$~dex found at higher halo masses.}
    \label{fig:stellarmass_halomass}
\end{figure}

We further examine the stellar-to-halo mass (SMHM) relation by jointly considering the stellar masses derived in Section~\ref{sn_stellarmass} and the GC-based halo masses described above. In Fig.~\ref{fig:stellarmass_halomass}, we present the distribution of UDGs in the $M_\star$--$M_{\rm h}$ plane, and compare it with the SMHM relation of \citet{rodriguez2017constraining}, extrapolated to lower stellar masses after correcting for the sign error in their Eq.~(66) (\citealt{rodriguez2017constraining}). This relation is based on a semi-empirical sub-halo abundance-matching approach and is broadly consistent with independent constraints from weak gravitational lensing, although the intrinsic scatter increases toward the low-mass regime relevant for UDGs ($\gtrsim 0.2$~dex).

Relative to this reference relation, we find that the majority of UDGs in our sample lie systematically above the expected SMHM locus at given stellar mass. In particular, $\sim 80\%$ of the UDGs exhibit halo masses higher than predicted for their stellar masses, often exceeding the upper envelope of the relation. This offset is especially pronounced for cluster UDGs, whereas field UDGs show a broader distribution, with some systems approaching the dwarf-galaxy locus. At the same time, the spread observed among cluster UDGs suggests that not all of them have undergone the same degree of environmental evolution. Part of this diversity may reflect differences in infall time and orbital history: UDGs accreted only recently may not yet have experienced sufficiently long-lived cluster processing to move far from the field-like locus, while those accreted earlier are more likely to occupy the most offset region of the SMHM plane. Such behavior closely mirrors that observed for Coma LSB galaxies and early-type dwarfs, while local dwarf galaxies---many of which are satellites subject to tidal stripping---tend to lie below the relation with comparatively lower halo masses.

An important caveat, when interpreting the scatter in the GC-based halo masses, is the uncertainty associated with GC candidate selection. The Euclid study of dwarf galaxies in the Perseus cluster by \citet{saifollahi2025euclid1}, for example, identified GC candidates using source morphology in the $I_E$ band together with the Euclid $I_E-Y_E$, $Y_E-J_E$, and $J_E-H_E$ colors. Although this combination provides valuable constraints on GC candidates, the absence of additional bluer ground-based optical photometry limits the color baseline available for separating genuine GCs from foreground stars and compact background galaxies. Such optical bands are particularly important because they provide complementary color information to the Euclid near-infrared data and can substantially improve the rejection of contaminants. Therefore, part of the relatively large scatter in the inferred GC abundances, and consequently in GC-based halo masses, may arise from uncertainties in candidate selection and contamination corrections, in addition to intrinsic galaxy-to-galaxy variation. This limitation highlights the importance of combining deep ground-based optical imaging with near-infrared observations for GC identification. Previous studies have shown that extending the wavelength baseline with optical--NIR colors can improve the separation of GCs from contaminants \citep{munoz2014next,voggel2024euclid,urbano2025euclid}. More generally, combining high-quality optical and near-infrared photometry should reduce the observational component of the scatter in GC-based halo-mass estimates, although intrinsic scatter and uncertainties in the $N_{\rm GC}$--$M_{\rm h}$ calibration will remain.

The systematic elevation of UDGs in the SMHM plane implies that, at a given stellar mass, UDGs typically reside in more massive dark-matter haloes than normal dwarf galaxies. This provides strong evidence that UDGs are highly dark-matter-dominated systems, characterized by low star-formation efficiency relative to their halo mass. A natural interpretation is that these galaxies formed their globular clusters efficiently at early times, when their haloes were rapidly assembling, but subsequently experienced suppressed or prematurely quenched field-star formation. As a result, stellar mass growth was curtailed, while the underlying dark-matter halo and the GC population were largely preserved. Our results therefore support a scenario in which UDGs represent an extension of the dwarf-galaxy population toward unusually high halo-to-stellar mass ratios, with the most extreme cases preferentially found in dense environments. However, this environmental trend should not be interpreted solely as a function of present-day location: it likely also encodes when UDGs entered the cluster, how long they have been exposed to cluster-specific processes, and whether they survived long enough to remain observable as UDGs at the present epoch.

\section{Conclusions}
\label{conclusion}

We have presented the first homogeneous and wide-area analysis of globular cluster (GC) systems associated with UDGs identified in the KiDS survey. By applying a dedicated GC-detection pipeline uniformly to a high-quality UDG catalog, we identified and characterized the GC populations of 24 GC-rich UDGs, including 14 systems associated with galaxy groups or clusters and 10 field systems, spanning environments from low-density regions to galaxy groups and clusters.

The pipeline was validated through direct comparisons with well-studied UDGs in the NGC~5846 group and the Fornax cluster. The recovered GC numbers and photometric properties are consistent with previous measurements within the quoted uncertainties, demonstrating the reliability of our GC detections and derived quantities despite the intrinsic depth limitations of the ground-based KiDS imaging.

Our analysis indicates a possible environmental dependence of GC richness in UDGs, although the effect appears to be weak. Cluster-associated UDGs tend, on average, to host somewhat larger GC populations and higher specific frequencies than their field counterparts. At fixed luminosity and stellar mass, cluster UDGs more often occupy the upper envelope of the $S_N$--$M_V$ and $N_{\rm GC}$--$M_\star$ relations. Field UDGs, by contrast, are still encompassed by the overall scatter of the UDG population, but tend to lie toward its lower part, closer to the locus of classical dwarf galaxies. The overlap between the two populations suggests, however, that present-day environment alone cannot fully account for the observed diversity. Some cluster UDGs may be relatively recent arrivals, for which cluster-specific processing has not yet had sufficient time to imprint stronger differences, while some field UDGs may retain GC properties similar to those of cluster systems because of similar early assembly histories or prior environmental preprocessing. These trends are nevertheless modest and do not support a clear environmental separation.

Using empirical GC--halo mass scaling relations to convert GC numbers into halo-mass estimates, we find that most UDGs in our sample reside in dark-matter haloes more massive than expected from the canonical stellar-mass--halo-mass relation of dwarf galaxies. This offset is particularly pronounced for cluster UDGs, implying unusually high halo-to-stellar mass ratios. However, the spread in halo-mass estimates, especially among cluster systems, suggests that the environmental dependence may also be modulated by infall time, orbital history, and the strength of the local tidal field, with these factors potentially contributing to the presence of UDGs at the most extreme, halo-dominated end of the distribution. These results reinforce the interpretation that UDGs are not a distinct galaxy class, but rather represent the high--halo-mass, low--star-formation-efficiency extension of the dwarf galaxy population.

Taken together, our findings are consistent with formation scenarios in which UDGs formed their GC populations efficiently during early halo assembly, followed by suppressed or prematurely quenched field-star formation. Our results also suggest that environmental effects should be considered as an important ingredient in the formation and evolution of both field and cluster UDGs, although the present data do not support a strong statistical separation between the two populations. The substantial overlap between field and cluster UDGs in $S_N$ and virial mass suggests that many present-day cluster UDGs may still closely resemble their pre-infall counterparts, having experienced only limited transformation since entering denser environments. In this picture, the role of environment is not necessarily to enhance the GC populations of UDGs, but rather to influence their subsequent evolution through quenching, preprocessing and survival. Systems exposed to stronger environmental processing may lose GCs or be transformed to the point that they no longer appear as UDGs, while the cluster UDGs that remain observable today may preferentially be those that have survived with their diffuse structure largely intact. 

The KiDS data presented in this work provide a valuable, high-quality ground based optical data that can be combined with space-based NIR data, e.g. from Euclid, to improve the GC selections in our UDG catalog, and 
push the analysis to the faint end of the UDG GC luminosity function. This will allow us to place tighter constraints on their dark-matter haloes and evolutionary histories of these systems.

\begin{acknowledgements}
      We thank Teymoor Saifollahi for his valuable comments and suggestions. This study was supported by the National Natural Science Foundation of China (NSFC) grant no 12573109. Rui Li acknowledges the support by the China Manned Space Program (no.CMS-CSST-2025-A03) and the National Natural Science Foundation of China (No. 12203050). SH acknowledges the support of the China Scholarship Council (grant n. 202406220048). 
\end{acknowledgements}

\bibliography{reference}{}
\bibliographystyle{aa}

\end{document}